\documentclass[12pt]{article}

\usepackage{scicite}
\usepackage{times}
\usepackage{graphicx} 
\usepackage{float}

\newenvironment{sciabstract}{%
\begin{quote} \bf}
{\end{quote}}

\usepackage{xcolor}
 
\usepackage{ulem}
 
\title{Attosecond-resolved photoionization of chiral molecules} 

\author
{S. Beaulieu$^{1,2^\ast}$, A. Comby$^{1}$,  A. Clergerie$^{1}$, J. Caillat$^{3}$, D. Descamps$^{1}$,\\
N. Dudovich$^{4}$, B. Fabre$^{1}$, R. G\'eneaux$^{5}$, F. L\'egar\'e$^{2}$, S. Petit$^{1}$, B. Pons$^{1}$,\\ 
G. Porat$^{4}$, T. Ruchon$^{5}$, R. Ta\"ieb$^{3}$, V. Blanchet$^{1}$  \& Y. Mairesse$^{1}$\\
\\
\normalsize{$^{1}$ Universit\'e de Bordeaux - CNRS - CEA, CELIA, UMR5107, F33405 Talence, France,}\\
\normalsize{$^{2}$ Institut Natinal de la Recherche Scientifique, Varennes, Quebec, Canada,}\\
\normalsize{$^{3}$ Sorbonne Universit\'es, UPMC Univ. Paris 6, CNRS-UMR 7614, LCPMR, 75252 Paris, France,}\\
\normalsize{$^{4}$ Department of Physics of Complex Systems, Weizmann Institute of Science, 76100, Rehovot, Israel,}\\
\normalsize{$^{5}$ LIDYL, CEA, CNRS, Universit\'e Paris-Saclay, CEA Saclay, 91191 Gif-sur-Yvette, France}\\
\\
\normalsize{$^\ast$To whom correspondence should be addressed; E-mail: beaulieus@emt.inrs.ca}
}

\date{\today}
\begin{document} 
\baselineskip24pt
\maketitle 

\begin{sciabstract}

Chiral light-matter interactions have been investigated for two centuries, leading to the discovery of many chiroptical processes used for discrimination of enantiomers. Whereas most chiroptical effects result from a response of bound electrons, photoionization can produce much stronger chiral signals that manifest as asymmetries in the angular distribution of the photoelectrons along the light propagation axis. Here we implement a self-referenced attosecond photoelectron interferometry to measure the temporal profile of the forward and backward electron wavepackets emitted upon photoionization of camphor by circularly polarized laser pulses. We found a delay between electrons ejected forward and backward, which depends on the ejection angle and reaches 24 attoseconds.  The asymmetric temporal shape of electron wavepackets emitted through an autoionizing state further reveals the chiral character of strongly-correlated electronic dynamics. 

\end{sciabstract}

Bolts and nuts are amongst the most common chiral objects in our macroscopic world. Their chiral nature is used to convert rotation to directional translation: rotating the nut on a bolt induces its translation forward or backward, depending on the rotation direction. A very similar effect occurs in the microscopic world when enantiopure chiral molecules are photoionized by circularly polarized radiation \cite{powis08}. The ejected photoelectrons tend to go forward or backward relative to the light propagation axis, depending on the helicity of the ionizing light and the handedness of the molecules \cite{ritchie76,bowering01}. As a result, the photoelectron angular distribution shows an asymmetry, called photoelectron circular dichroism (PECD). PECD is one of the most sensitive probes of static \cite{nahon15} and dynamical \cite{comby16,beaulieu16-1} molecular chirality, producing signals that are up to two orders of magnitude larger than most circular dichroisms. From a classical point of view, PECD can be seen as the result of the combined action of the chiral molecular potential and the circular ionizing electric field on the outgoing electron trajectories \cite{beaulieu16}. Quantum mechanically, it arises from the interference between partial waves of different parity constituting the outgoing photoelectron wavepacket \cite{powis08}. Both interpretations show that subtle differences in the ionization dynamics can have dramatic consequences. Consequently, PECD has been proposed as a possible hypothesis to explain the homochirality of terrestrial life \cite{meierhenrich08}: the asymmetric electron ejection induces an asymmetric recoil of the ions, which can lead to enantiomeric separation when accumulated over hundreds of millions of years \cite{tia13}.

\begin{figure}
\begin{center}
\includegraphics[width=0.8\textwidth]{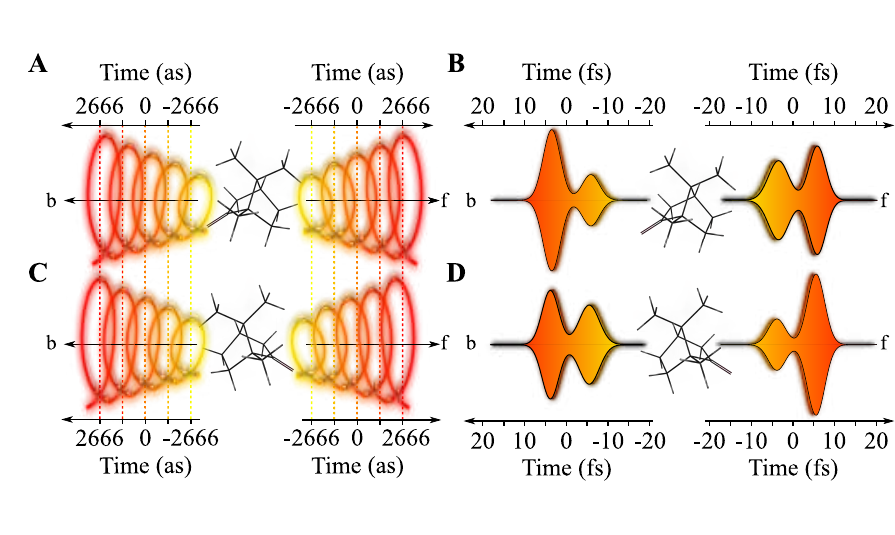}
\caption{Schematic view of the two timescales of photoionization of two camphor enantiomers. (1S)-(-)-camphor is shown in (\textbf{A}) and (\textbf{B}), (1R)-(+)-camphor in (\textbf{C}) and (\textbf{D}). In direct photoionization (A), (C) the forward (f) and backward (b) electron wavepackets may be delayed by a few attoseconds because of the asymmetric scattering  of the outgoing electron in the chiral molecular potential. In the case of autoionization (B),(D) the dynamics of the autoionizing state can lead to different temporal structures of photoelectron wavepackets in the forward and backward directions, on the femtosecond timescale.}
\label{fig0}
\end{center}
\end{figure}

Photoionization was considered to be instantaneous from an experimental point of view until attosecond technology made it possible to measure the underlying ultrafast electron dynamics. Delays of a few attoseconds were measured between electrons originating from different atomic orbitals \cite{klunder11}, from distinct bands of a solid \cite{cavalieri07}, associated with different vibrational states of a molecular ion \cite{haessler09} or from differing spin-orbit states \cite{jordan17}. The direction of the electron emission also influences the photoionization dynamics: delays have been observed between electrons ejected at different angles \cite{heuser16,hockett16,baykusheva17} or from different sides of an asymmetric molecule \cite{chacon14,cattaneo16-1}.

The photoionization process involves more complicated dynamics when autoionization occurs. In that case, the photoabsorption promotes the system into a metastable bound state which is coupled to equienergetic continuum states through configuration interaction. This coupling leads to autoionization of the metastable state. The interference of direct and indirect photoionization channels produces characteristic (Fano) spectral profiles \cite{fano61}, associated with complex temporal dynamics. Pioneering attosecond experiments in rare gases recently showed the possibility of measuring the buildup of Fano lineshapes in the temporal domain \cite{kaldun16} as well as the spectral phase across the resonance \cite{kotur16,gruson16}, allowing the reconstruction of the temporal profile of the electron wavepacket \cite{gruson16}. 

\begin{figure}
\begin{center}
\includegraphics[width=0.8\textwidth]{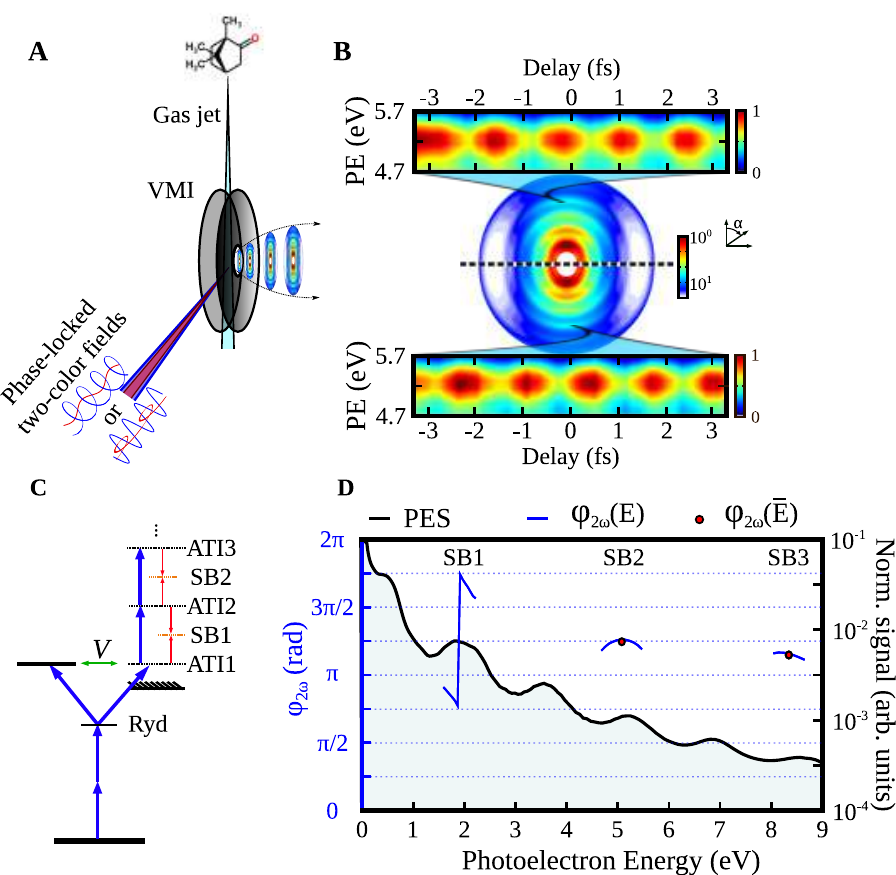}
\caption{Principle and implementation of photoelectron interferometry. (\textbf{A}) Schematic experimental setup. Two phase-locked femtosecond laser pulses with linear or circular polarization are focused into a jet of enantiopure camphor molecules in the interaction zone of a velocity map imaging (VMI) spectrometer. The photoelectrons are accelerated by a set of electrodes onto microchannel plates imaged by a phosphor screen and a CCD camera, which records the 2D projection of the 3D angular distribution of the photoelectron spectrum with an energy resolution around 0.2 eV at 2 eV. (\textbf{C}) Ionization scheme. The 400 nm pulse (40 fs duration, intensity $\sim 5\times 10^{12}$ W/cm$^2$) induces multiphoton above-threshold ionization of the molecules. The first ionizing transition lies in the vicinity of an autoionizing resonance. The 800 nm pulse (30 fs, intensity $\sim 5\times 10^{11}$ W/cm$^2$) produces additional transitions, leading to the creation of sidebands between the ATI peaks. (\textbf{B}) Typical measured photoelectron angular distribution and evolution of the second sideband as a function of delay between the two fields. The oscillations in the upper and lower half of the distribution are out of phase, reflecting the up-down asymmetry of the total ionizing electric field. (\textbf{D}) Angle-integrated photoelectron spectrum, constituted of ATI peaks and n-th order sidebands (SBn). The full blue lines are the angularly-integrated spectrally-resolved $2\omega$-oscillation phases for each sideband and the red dots are the spectrally-averaged $2\omega$-oscillation phases for non-resonant SB2 and SB3.}
\label{fig1}
\end{center}
\end{figure}

Here, we aimed to determine whether or not the electrons ejected forward and backward from a sample of randomly oriented enantiopure chiral molecules were temporally synchronized. Answering this question, for both direct (Fig. \ref{fig0}(A,C)) and indirect (Fig. \ref{fig0}(B,D)) photoionization pathways, is a challenging task. Up to now, attosecond delay measurements in the gas phase have been conducted on rare gas atoms or di- and triatomic molecules, which were used as benchmark systems. These experiments have revealed a strong influence of the weak probe field on the outcome of the measurement. It is thus necessary to perform accurate theoretical calculations to calibrate these measurement-induced delays \cite{klunder11,ossiander17}. Such theoretical calculations for large and low-symmetry chiral molecules (\textit{e.g.} camphor, C$_{10}$H$_{16}$O) are currently far from reach. 

In order to directly access the delays between forward and backward electron emission, without any measurement-induced effects, we implemented a self-referenced photoelectron interferometry technique using photoionization by two phase-locked laser fields to detect differential attosecond photoionization delays with a resolution of 2 as. By independently controlling the chirality of the ionizing and probe light pulses, we fully decoupled the intrinsic photoionization delays from the measurement-induced delays. 

\paragraph*{Photoelectron interferometry}

When an intense femtosecond laser pulse ionizes an atom or a molecule, multiple photons can be absorbed above the ionization threshold (Above-Threshold Ionization, ATI). In the spectral domain, ATI produces a comb of photoelectron peaks separated by the laser photon energy \cite{agostini79}. Each peak is characterized by a spectral width $\delta \omega$, and the overall ATI spectrum extends over a width $\Delta \omega$. In the time domain, the ATI process leads to the emission of attosecond electron bursts (Fig. \ref{fig0}(A-C)) of duration $\delta t$, which form a train. The overall duration of the train $\Delta t$ is set by the laser pulse duration, typically a few tens of femtoseconds. The ATI emission can last longer if an autoionizing state is populated: the lifetime of the autoionizing state increases the electron wavepacket duration (Fig. \ref{fig0}(B-D)). Characterizing the temporal dynamics of the ionization process requires measuring the process on two timescales $\delta t$ and $\Delta t$. The femtosecond structure of the wavepacket is encoded in the spectral intensity and phase within the bandwidth of each ATI peak: $\Delta t$ is related to $\delta \omega$.  On the other hand, the attosecond sub-structures are encoded in the relative amplitude and phase between the different ATI peaks: $\delta t$ is associated to $\Delta \omega$. In order to have a complete picture of the temporal dynamics of the ionization process, it is thus necessary to measure the spectral phase of the ATI peaks, both within their bandwidth ($\delta \omega$) and from one peak to the next ($\Delta \omega$). This is possible using photoelectron interferometry \cite{paul01-1,zipp14,gruson16}.
 
We first present the basic concepts of photoelectron interferometry and highlight the rich spectroscopic information that it provides about the ionized target. For now, we are leaving aside the chiral character of the experiment (by integrating over the photoelectron ejection angles). The principle of the measurement is described in Fig. \ref{fig1}. The target molecule we chose was camphor, a bicyclic ketone which has been extensively studied in PECD experiments performed in single-photon \cite{powis08,nahon16}, in multiphoton \cite{lux12,lehmann13} and in ATI regimes \cite{lux15}. Camphor has a first ionization potential of 8.76 eV and Rydberg states starting around 6.2 eV: it is ionized by $2+n$ resonance-enhanced multiphoton transitions when using a 400 nm ultraviolet (UV) field, with $n$ being the order of the ATI peak. In the presence of a weak infrared (IR) 800 nm field (frequency $\omega$), new peaks, called sidebands, appear between the ATI comb. Two quantum paths lead to the same sideband: addition of an IR photon to ATI peak $n$, or subtraction of one IR photon from ATI peak $n+1$. These two paths interfere and the sideband amplitude oscillates as a function of the relative delay between the UV and IR fields, at $2\omega$ frequency \cite{schumacher94,zipp14,gong17}. The phase of the sideband oscillations encodes the relative phase between the two neighboring ATI peaks and thus the temporal properties of the photoemission process. 

The photoelectrons were collected by a velocity map imaging spectrometer (VMI), which measures the angle-resolved photoelectron spectrum (Fig. \ref{fig1}(A,B)). The superposition of 800 nm and 400 nm pulses produces an electric field which is stronger in the upper or lower direction, depending on the relative delay between the two fields. As a consequence, the electrons ejected up and down are modulated in opposite phase \cite{zipp14,skruszewicz15} (Fig. \ref{fig1}(B)). We measured the sideband phase independently on the angularly-integrated upper and lower half of the photoelectron image and averaged the phase obtained from the upper half with the  $\pi$-shifted phase obtained from the lower half. More details about the experimental setup, raw VMI images and their inversion \cite{garcia04} as well as data analysis are given in Supplementary Material (SM). 

Figure \ref{fig1}(D) shows the sideband oscillation phase $\varphi_{2\omega}$ as a function of the photoelectron kinetic energy $E$. The first sideband (SB1), which encodes the phase difference between ATI peaks 1-2, presents an abrupt $\pi$-phase jump around 1.9 eV. This is the signature of a resonance associated with one or the other of the two contributing ATI peaks. The second and third sidebands, which are built respectively upon ATI peaks 2-3 and 3-4, have a smooth phase variation across their bandwidth (\textit{i.e.} without any trace of resonance). We can thus conclude that the resonance occurs in the formation of the first ATI peak (ATI1), and does not propagate to the higher ATI peaks, as confirmed by theoretical calculations of resonant photoelectron interferometry presented in the SM. 

We proceeded to investigate the chiral (enantio-specific) photoionization dynamics in the two different regimes identified above: direct -- attosecond -- ionization (SB2 and SB3), and indirect -- femtosecond -- ionization in the vicinity of an autoionizing resonance (SB1). 

\paragraph*{Attosecond delays in non-resonant photoionization}

We started by analyzing the direct photoionization dynamics which occur on the attosecond timescale and are encoded in the relative phase between the different ATI peaks. It can be obtained by extracting the oscillation phases of the signals averaged over the bandwidth of each sideband, $\varphi_{2\omega}(\overline{E})$. Neglecting the variations over the spectral width is equivalent to assuming that the photoionization process is strictly periodic from one laser cycle to the next. The spectral homogeneity of the sideband phases shown in Fig. \ref{fig1}(D) indicates that this assumption is reasonable for SB2 and SB3, but not for the resonant SB1, which is discussed later. This scheme is similar to the conventional RABBIT analysis (Reconstruction of Attosecond Beatings By Interference of Two-photon transitions) \cite{paul01-1}, here extended to the case of multiphoton ionization \cite{zipp14,gong17}. Simulations presented in the SM validate the analogy between the two techniques.   

\begin{figure}
\begin{center}
\includegraphics[width=0.8\textwidth]{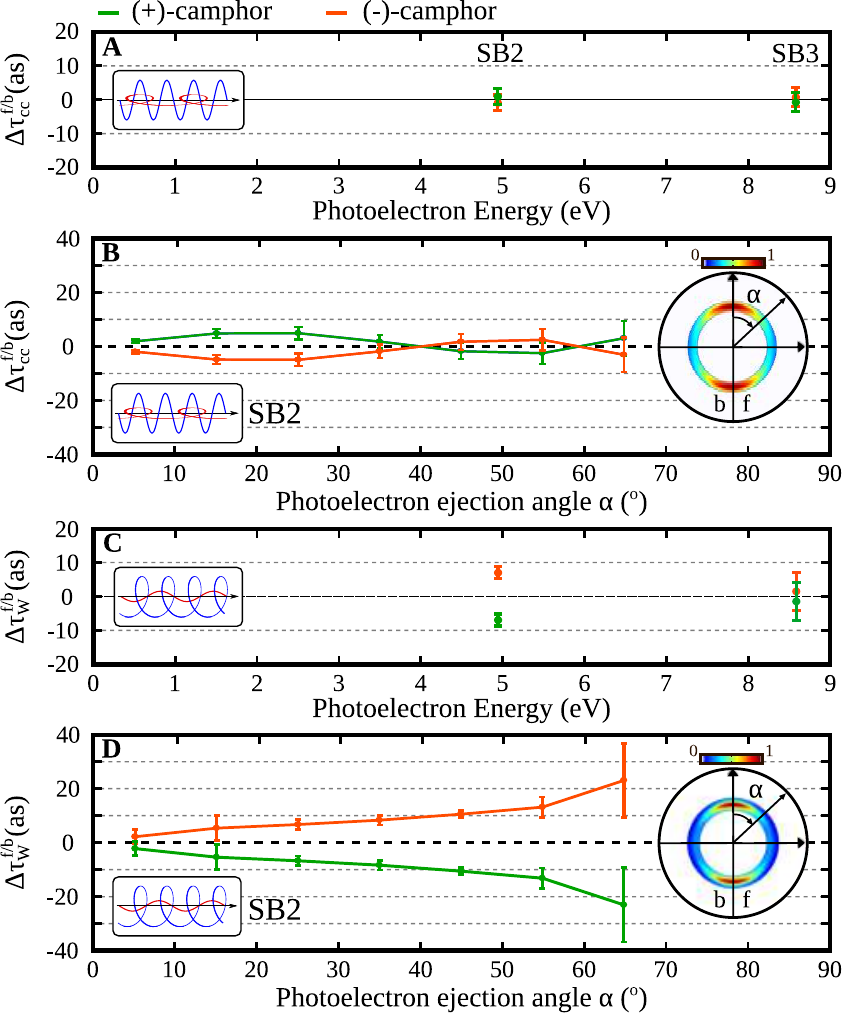}
\caption{Forward/backward differential delays in non-resonant photoionization of camphor. (\textbf{A}) When the UV is linearly polarized and the IR is left-circularly polarized, the differential delay $ \Delta \tau^{f/b}_{cc}$ is zero for SB2 and SB3. (\textbf{B}) Resolving angularly the differential delay shows that it reaches non-zero values for electrons ejected close to the laser polarization plane (for SB2). The inset depicts the angular distribution of the normalized SB2 intensity. (\textbf{C}) When the UV field is circularly polarized and the IR one is linearly polarized, the differential delay $ \Delta \tau^{f/b}_{W}$ is non zero for SB2. (D) Angular resolution of the differential delays from SB2.  Error bars are defined as the 95$\%$ confidence interval.}
\label{fig2}
\end{center}
\end{figure}

Measuring the phase $\varphi_{2\omega}(\overline{E})$ is equivalent to measuring the time delay $\tau$ that maximizes the signal of each sideband. This delay can be decomposed as the sum of three contributions, that reflect the three steps of the sideband creation \cite{dahlstrom12}: 

\begin{equation}
 \tau = \tau_{light}  + \tau_{W} + \tau_{cc} 
 \label{eq}
\end{equation}

The ionization is triggered by absorption of light at a well defined time $\tau_{light}$. Next, the electron scatters in the molecular potential and acquires a delay $\tau_{W}$ -- the Wigner delay \cite{wigner55}. The electron also interacts with the weak IR field, which induces continuum-continuum transitions from the main ATI peaks to the sidebands, introducing an additional delay $\tau_{cc}$. Whereas $\tau_{light}$ and $\tau_{cc}$ are induced by the measurement, $\tau_{W}$ is intrinsic to the probed system and is the physical quantity of interest. It represents the delay between an electron scattering in a given potential and in a reference potential, as introduced by Wigner in 1955 \cite{wigner55}. 

To resolve the enantiosensitivity of Wigner delays, we turned to chiroptical measurements comparing the sideband oscillation phases for electrons emitted in the forward vs. backward directions, and extracting the difference $\Delta \tau^{f/b} = \tau^{f} - \tau^{b} $. This procedure naturally eliminates $\tau_{light}$, which is strictly common to the forward and backward electrons. Further decoupling is achieved by using different combinations of linearly and circularly polarized light. Indeed, the forward/backward (f/b) symmetry can only be broken by the chiral nature of the interaction, that is, if a circularly polarized light pulse is used. We can thus selectively break the f/b symmetry only in the ionization step by using a circularly polarized UV field and a linear IR field. In that case $\Delta \tau^{f/b}_{cc}$ = 0 and $\Delta \tau^{f/b}=\Delta \tau^{f/b}_{W}$. Alternatively, we can render the Wigner delay f/b symmetric by using linearly polarized UV for ionization and circularly polarized IR probing photons to obtain $\Delta \tau^{f/b}_{W}=0$ and $\Delta \tau^{f/b}=\Delta \tau^{f/b}_{cc}$. 

The photoelectron images were separated in four quadrants, and the signal was angularly-averaged in each quadrant. A Fourier analysis of the $2\omega$-oscillations was conducted to determine the delay that maximized each sideband, in each quadrant. We calculated the difference between the delays measured in the forward and backward directions $\Delta \tau^{f/b}$. This procedure was repeated for left and right helicities, and for five consecutive measurements of each enantiomer. The delays measured from opposite helicities or opposite enantiomers have opposite signs, revealing that the differential f/b ionization delay is a genuine chiral observable. In order to extract the most accurate value of the differential delay, we averaged the results obtained from the (+) and (-) enantiomers: $\Delta \tau^{f/b}=(\Delta \tau^{f/b}_{(+)} -\Delta \tau^{f/b}_{(-)})/2$ (see SM). 

First, we used a linearly polarized UV ionizing field and a circularly polarized IR measurement field. The Wigner component $\tau_W$ of the sideband delay was thus f/b symmetric ($\Delta \tau^{f/b}=\Delta \tau^{f/b}_{cc}$). The results shown in Fig. \ref{fig2}(A) reveal that the differential f/b continuum-continuum induced delay is zero (within the 2 as accuracy of the present measurement). This means that the laser-induced transitions produce essentially the same delay on electrons emitted in the forward and backward directions, without any significant sensitivity to the chiral character of the ionic potential. However, a weak influence of the chiral potential is still noticeable in this polarization configuration: the intensity of SB2 averaged over all UV-IR delays, shows a f/b asymmetry (PECD) on the order of 0.5$\%$ (twice smaller than when the UV is circularly polarized). 

In order to find signatures of chirality in the photoionization delays, we resolved the angular dependence of the photoionization dynamics \cite{aseyev03,heuser16,hockett17}. We integrated the photoelectron signal in slices of 10$^\circ$ around different ejection angles $\alpha$ from the polarization plane of the light, and measured the associated delays ($\Delta \tau^{f/b}_{cc}$). The results for SB2 are shown in Fig. \ref{fig2}(B).  For electrons emitted beyond 70$^\circ$, the signal was too low to extract reliable values. A weak but non-zero $\Delta \tau^{f/b}_{cc}$ is measured when electrons are ejected close to the polarization plane of the IR laser, reaching 5 $\pm 2$ as at $\alpha=25^\circ$. This delay tends to vanish for higher ejection angles but the error bars become larger due to the lower level of signal. Measurements on SB3 show a zero delay whatever the ejection angle (see SM). In the commonly accepted intuitive picture of multicolor photoionization, the linear UV field induces bound-free transitions starting from the molecular core region, and the IR field subsequently drives continuum-continuum transitions while the electron is escaping from the core region. The continuum-continuum transitions are thus rather insensitive to the details of the molecular potential. This interpretation was recently confirmed by the observation of a zero continuum-continuum delay between electrons escaping the two sides of an oriented asymmetric molecule in theoretical calculations \cite{chacon14}. The zero overall delay we measured between forward and backward electrons on SB2 and SB3 is consistent with this picture. Nevertheless, the angle-resolved measurements demonstrate that the continuum-continuum transitions can be slightly influenced by the core (chiral) region of the potential. 

Second, we broke the f/b symmetry in the ionization step by switching the polarization state of the ionizing UV field to circular while using a linearly polarized IR field. The photoionization process is here intrinsically f/b asymmetric, while the continuum-continuum coupling is f/b symmetric: $\Delta \tau^{f/b} = \Delta \tau^{f/b}_{W}$. The magnitude of the temporally-averaged PECD on the SB2 was larger than in the previous configuration, reaching 1$\%$. The measurements (Fig. \ref{fig2}(C)) show a differential delay of $\Delta \tau^{f/b}_{W} = 7 \pm 2$ attoseconds for the second sideband (SB2). Our experiment is thus able to reveal a small f/b asymmetry in the Wigner delay in the photoionization of chiral molecules by circularly polarized light. The $\Delta \tau^{f/b}$ vanishes for SB3 because of the decrease of both the f/b asymmetric character of the photoionization and the absolute Wigner time delay with increasing photoelectron kinetic energy.

The evolution of the differential Wigner delays with photoelectron ejection angle are shown in Fig. \ref{fig2}(D). Close to $\alpha$ = 0$^\circ$, $\Delta \tau^{f/b}_{W}$ is null, which is not surprising since the PECD also vanishes in the laser polarization plane. For electrons emitted in the 60-70$^\circ$ slice,  $\Delta \tau^{f/b}_{W}$ reaches 24 as. This angle-resolved analysis shows that while the average difference between forward and backward electron ionization times is only 7 as, it strongly varies with ejection angle and can reach higher values for electrons emitted away from the laser polarization plane. Repeating this analysis for SB3 shows that the differential Wigner time delay remains zero within the error bars for all photoelectron ejection angles (see SM). By accessing the angular dependence of the emission time, our measurements give access to the phase properties of the photoionization matrix elements. The determination of the underlying scattering phase shifts has been a long-standing quest of photoionization experiments, and our results show that their energy-derivative, \textit{i.e.} the Wigner delays, are accessible with high accuracy using a relatively simple setup. 

The differential Wigner delay is a signature of the asymmetric scattering process which is at the heart of the photoelectron circular dichroism phenomenon. Wigner delays are determined by the energy derivative of the scattering phase. We performed a theoretical analysis of the photoionization of camphor molecules (see SM for details about the theoretical model). Our calculations confirm the existence of asymmetric Wigner delays, even in a randomly oriented ensemble of molecules. The theoretical forward/backward differential Wigner delay $\Delta \tau^{f/b}_{W}$ is of the order of 5 as for 2 eV electrons, which agrees with the present experimental observation. Interestingly, the evolution of $\Delta \tau^{f/b}_{W}$ with respect to the photoelectron energy, shown in the SM, shows rich spectroscopic features that are not visible in the photoelectron spectrum and cannot be easily distinguished in the PECD signal. Thus $\Delta \tau^{f/b}_{W}$ is a remarkable chiral observable, which enables tracking of subtle features of the molecular potential such as, for example, the differential Cooper minima, surviving the molecular orientation averaging (see SM). This observation opens prospects for highly sensitive experiments, for instance through molecular-frame measurements \cite{tia17}, as well as accurate testing of advanced quantum theories of molecular photoionization.

The differential f/b analysis and the control of the chiral symmetry breaking of the interaction enabled us to fully decouple the different components of the photoionization delays, without the need for any theoretical input, and to reveal a tiny but measurable delay in the direct photoionization. We proceeded to use photoelectron interferometry to extract more complex dynamics occurring when the chiral molecules are photoionized in the vicinity of an autoionizing resonance. 

\paragraph*{Resonant photoionization}
Continuum resonances play an essential role in the photoionization of most polyatomic molecules. They can arise from the shape of the molecular potential, in a single-electron picture (shape resonances) \cite{piancastelli99}, or from multi-electron dynamics involving electron correlations and couplings between different channels \cite{fano61}. In both cases, spectrally localized scattering phase jump(s) are expected, reflecting the modification of the ionization dynamics. For instance, in chiral molecules, the PECD was recently shown to be enhanced in the vicinity of a Fano resonance \cite{catone12}. Here we used the photoelectron interferometry technique to directly track the asymmetric ionization dynamics of chiral molecules in the vicinity of an autoionization resonance. 

\begin{figure}
\begin{center}
\includegraphics[width=0.8\textwidth]{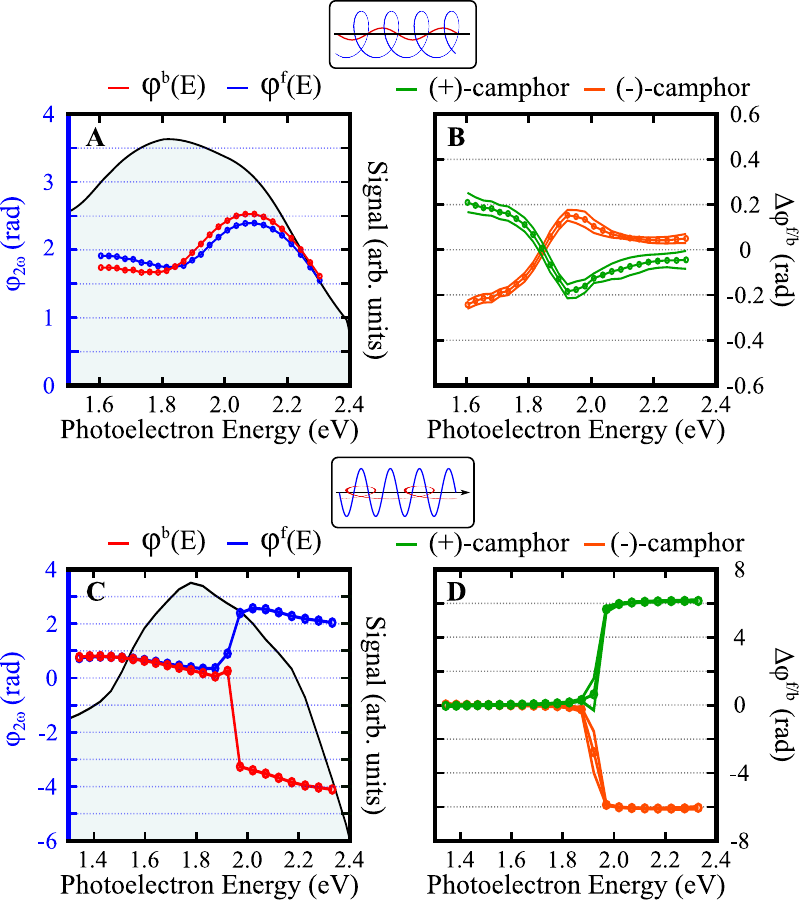}
\caption{Phase-resolved resonant photoionization in camphor.  (\textbf{A}),(\textbf{C}) Spectral amplitudes (black) and forward and backward spectral phases ($\varphi^{f}(E)$, blue and $\varphi^{b}(E)$, red) of SB1 in (1R)-(+)-camphor, using left-circularly polarized UV - linearly polarized IR (A) and linearly polarized UV - left-circularly polarized IR (C). (\textbf{B}),(\textbf{D}) forward backward asymmetry of the spectral phase ($\Delta \varphi^{f/b}$) in camphor, using left-circularly polarized UV - linearly polarized IR in (B) and linearly polarized UV - left-circularly polarized IR (D). In (B) and (D), the dots represent the mean values of the forward backward asymmetry of the spectral phase while the solid lines show the error bars, which are defined as the 95$\%$ confidence interval.}
\label{fig3}
\end{center}
\end{figure}

As shown in Fig. \ref{fig1}(D), the SB1 presents a sharp $\pi$-phase jump around 1.9 eV, which reflects the presence of a resonance on the first ATI peak. We resolved this phase jump in the forward (blue) $\varphi^F$ and backward (red) $\varphi^B$ directions, using circularly polarized UV (Fig. \ref{fig3}(A)) or IR (Fig. \ref{fig3}(C)) light. When the UV light is circularly polarized, the spectral phase exhibits a weak $\sim$ 0.75 rad bump centered around 2.1 eV. A significant difference was seen between the forward and backward spectral phases. This difference ($\Delta \varphi^{f/b} = \varphi^F - \varphi^B$) (Fig. \ref{fig3}(B)) shows a good mirroring between the two enantiomers.

The case where the f/b symmetry is broken by the weak IR pulse (linear UV and circular IR) is more intriguing. The spectral phases show a steep $\sim \pi$ jump around 1.9 eV, in opposite directions for forward and backward electrons. After the jump, the phases become nearly identical, as they are separated by $\sim 2 \pi$ (Fig. \ref{fig3}(C)). The f/b differential phases ($\Delta \varphi^{f/b}$) obtained in the two enantiomers are almost exactly opposite (Fig. \ref{fig3}(D)). To a first approximation, the presence of this huge asymmetry is unexpected. The circularly polarized field acts during the continuum-continuum transitions, which should be affected mostly by the long-range (non-chiral) part of the molecular potential and should, therefore, be f/b symmetric \cite{chacon14}. Our measurement demonstrates that in the vicinity of a resonance, the f/b symmetry can also be broken during the continuum-continuum transitions. This finding is in agreement with a recent theoretical investigation of photoelectron interferometry, which demonstrated that the simple separation of the measured delay ($\tau$) in a sum of the contributions from Wigner ($\tau_W$) and continuum-continuum ($\tau_{cc}$) delays did not hold anymore in the presence of a resonance \cite{argenti17}. Indeed, Argenti \textit{et al.} demonstrated that $\tau_W$ and $\tau_{cc}$ are entangled in resonant photoionization. The measured delay ($\tau$) is representative of the two-color photoionization process, and our results show that the circular polarization of the weak IR field is sufficient to induce a major symmetry breaking, in the presence of a resonance.

\begin{figure}
\begin{center}
\includegraphics[width=1\textwidth]{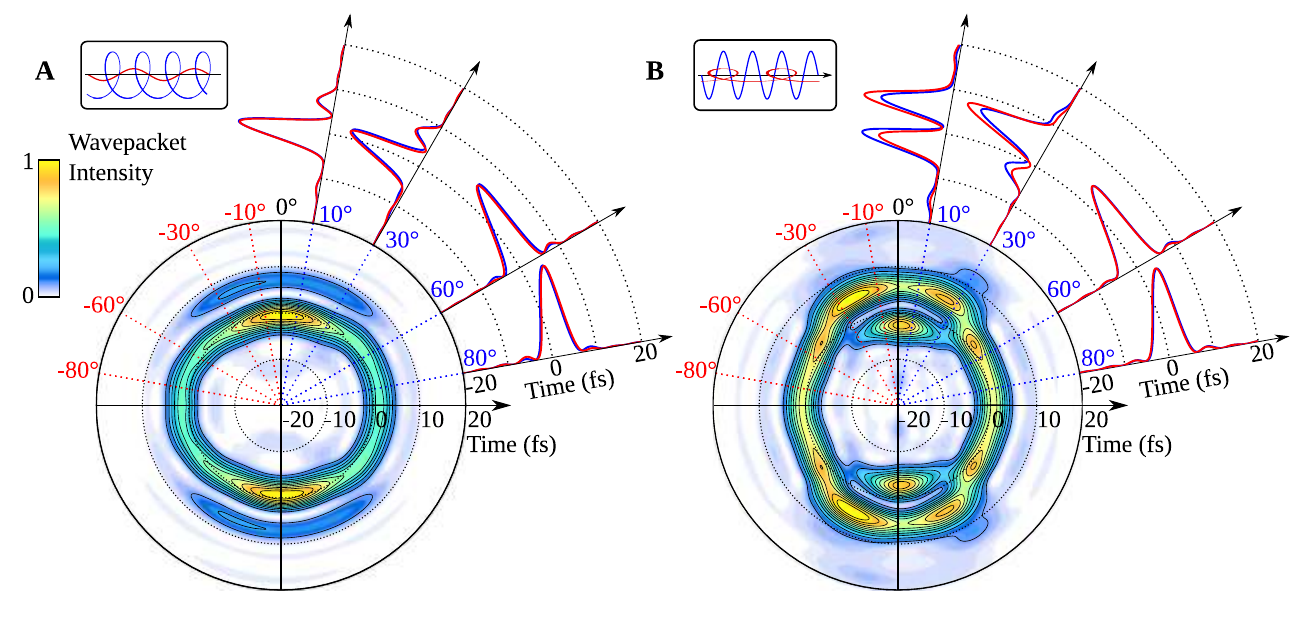}
\caption{Angle-resolved temporal profile of the autoionizing photoelectron wavepacket.  The contourplot depicts the temporal profile of the wavepacket as a function of the electron ejection angle. The upright external plots show forward (blue) and backward (red) cuts of the wavepackets along specific angles. (\textbf{A}): left-circularly polarized UV, linearly polarized IR. (\textbf{B}): linearly polarized UV, left-circularly polarized IR. }
\label{fig4}
\end{center}
\end{figure}

\paragraph*{Asymmetric electron wavepackets}
We could retrieve the temporal profiles of the two-color forward and backward wavepackets by Fourier-transforming their measured spectral amplitudes and phases. In order to extract the angular dependence of these temporal profiles, we analyzed the oscillations of SB1 as a function of the electron ejection angle ($\alpha$), as done previously in the non-resonant case (see raw data in SM). Figure \ref{fig4} shows the resulting angle-resolved wavepackets in the temporal domain. In both polarization configurations, the wavepacket shows a single temporal peak when the electrons are ejected close to the propagation axis of the light (90$^\circ$), and a double peak structures when the electrons are ejected near the laser polarization plane (0$^\circ$). The latter are signatures of the temporal interference between the direct non-resonant and the resonant components of the autoionizing wavepackets \cite{gruson16}. The effect of the resonance appears more confined around the laser polarization direction when the UV field is linearly polarized (Fig. \ref{fig4}(B)), probably because of a stronger anisotropy of the resonant excitation compared to the circularly polarized case. 

The chiral nature of the photoionization process can be investigated by comparing cuts of the temporal profile of electrons ejected at positive (forward, blue) and negative (backward, red) angles ($\alpha$). When the UV field is circularly polarized, the two bumps of the forward electron wavepacket emitted around 30$^\circ$ maximize $\sim 400$ as after the backward wavepacket (Fig. \ref{fig4}(A)). Interestingly, a similar delay is measured around 60$^\circ$, where the wavepacket shows a single peak structure. As the ejection angle further increases, the ordering between forward and backward emission reverses, with a $\sim - 250$ as delay around 80$^\circ$. These subtle features obey chiral inversion when switching from one enatiomer to the other, as shown in the SM. In the other polarization configuration (Fig. \ref{fig4}(B)), the two bumps from the forward and backward wavepackets are synchronized in time for electrons ejected close to the laser polarization plane ($\alpha$ = 0$^\circ$). However their relative yield is strongly f/b asymmetric. This means that in the vicinity of resonances, where $\tau_{W}$ and $\tau_{cc}$ are entangled \cite{argenti17}, the perturbative IR pulse can be used to break the f/b symmetry and to subsequently tailor asymmetric electronic wavepackets, both in time and space.  At larger emission angles, where the dynamics are governed by a single -non-resonant- pathway, the single-peak wavepackets become forward-backward symmetric. This analysis provides a deep insight into the angular-dependence of the multielectron dynamics governing autoionization. 

Comparing the two polarization configurations used in the measurements (Fig. \ref{fig4}) shows that the wavepacket asymmetry is in fact much stronger when it is the weak IR field that is circularly polarized. We attribute this result to the sequential nature of the resonant photoionization process. The linearly polarized UV photons populate a quasi-bound state embedded in the continuum, which can relax by ionization, releasing electrons at the energy of the first ATI peak (autoionization, Fig. \ref{fig1} (C)). However another process could lead to ionization of the quasi-bound state: the absorption of one IR photon releasing an electron with the energy of the first sideband. This can be seen as a classic PECD experiment, starting from a highly excited quasi-bound state. Recent experiments showed that PECD could be observed when bound states excited by linear photons were ionized by circularly 
polarized photons \cite{comby16}. The present scheme extends this scenario to quasi-bound states.  On the other hand, when the UV photons are circularly polarized, they can induce an asymmetric wavepacket in the excited states, a phenomenon called PhotoeXcitation Circular Dichroism (PXCD) \cite{beaulieu16-2}. The ionization of such a wavepacket by linearly polarized light produces f/b asymmetries, but they were observed to be weaker than the PECD from excited states. This could explain why we observe a weaker wavepacket asymmetry when the IR photons are linearly polarized.

\begin{figure}
\begin{center}
\includegraphics[width=1\textwidth]{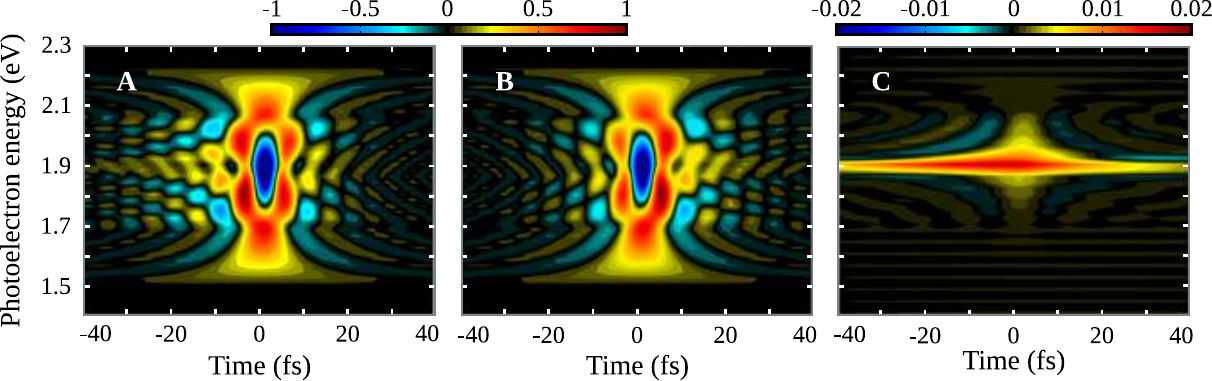}
\caption{Wigner-Ville distributions of the autoionizing photoelectron wavepackets. The distributions were calculated on an average wavepacket obtained by summing the wavepacket from (1R)-(+)-camphor and the mirrored wavepacket from (1S)-(-)-camphor. (\textbf{A}) shows the WVD of a forward wavepacket emitted around $\alpha$ = 10$^\circ$ from the linearly polarized UV field, and (\textbf{B}) the WVD of a backward wavepacket emitted around $\alpha$ = -10$^\circ$ (\textbf{C}) is the WVD of the forward-backward differential wavepacket.}
\label{fig5}
\end{center}
\end{figure}

\paragraph*{Time-frequency analysis}
The temporal profile of the wavepackets only provides spectrally integrated information about the rich ongoing dynamics. The spectral origin of temporal asymmetries can be revealed using a time-frequency analysis \cite{busto17}. The Wigner-Ville distribution (WVD) is particularly interesting because it encodes the quantum interference between different components of a wavepacket. For a wavefunction $\Psi(t)$, the WVD ($W(\Omega,t)$) is defined as: 

\begin{equation}
W(\Omega,t)=\int \Psi(t-\tau/2) \Psi^*(t+\tau/2) e^{i\Omega\tau} d\tau
\end{equation}

where $\Omega$ is the angular frequency, and $t$ is the time. Figure \ref{fig5} shows the WVD of the electron wavefunctions emitted around $\alpha$ = +10$^\circ$ (forward, $\Psi^f(t)$) and $\alpha$ = -10$^\circ$ (backward, $\Psi^b(t)$)) from the laser polarization plane. The distributions were calculated by averaging the wavepacket from (+)-camphor and the mirrored wavepacket from (-)-camphor. A strong negative lobe is present around time $t=0$ at the energy of the resonance (1.9 eV), revealing the quantum interference between the direct and indirect ionization components. Hyperbolic fringes converging to the energy of the resonance are observed in the leading or falling edge of the distribution, depending on the f/b emission direction. The asymmetry of these fringes reflects an asymmetric destructive interference between wavepacket components. In order to isolate the asymmetric part of the wavefunction, we calculate the WVD of the differential wavefunction $\Delta\Psi^{fb}(t)=\Psi^f(t)-\Psi^b(t)$ (Fig. \ref{fig5}(C)). They are strikingly simple, with a temporally long and spectrally narrow signal at the energy of the resonance, and almost no negative components. We conclude that while the forward and backward wavepackets are each formed by the coherent superposition of a resonant and a non-resonant contribution, the chiral character of the wavepacket appears to be strongly dominated by a single - resonant - pathway, which leads to the disappearance of the signature of quantum mechanical interference in the WVD. The WVD thus provides unique insight into the origin of the asymmetric shaping of photoelectron wavepackets during resonant photoionization of chiral molecules.

\paragraph*{Conclusions and Perspectives}
Our results show that using circularly polarized photons to drive photoionization of chiral molecules induces asymmetric delays in the photoemission, on both femtosecond and attosecond timescales. In direct non-resonant photoionization, the forward/backward asymmetry in the Wigner time delays are on the order of few attoseconds. In the vicinity of an autoionizing resonance driven by electronic correlation, we have found that the emitted two-color electron wavepacket is strongly asymmetric, demonstrating the chiral character of this multielectronic effect. By using the synergies of molecular and light chirality in the vicinity of resonances, we demonstrated tailoring of the shape of the released electron wavepackets, both in time and space, which is a new scheme for multidimensional attosecond quantum control. The high accuracy of the measurements can also be used as a powerful benchmarking tool for quantum theories of molecular photoionization. Finally, our approach could be generalized to a broad variety of systems to shed light on the ultrafast symmetry breakings which are at the heart of very recent technological and scientific breakthroughs, such as in superconducting chiral nanotubes \cite{qin17} and chiral spintronics devices \cite{naaman15}.  


\section*{Acknowledgments}

We thank R. Bouillaud and L. Merzeau for technical assistance, and Anne L'Huillier for fruitful discussions. This project has received funding from the European Research Council (ERC) under the European Unions Horizon 2020 research and innovation program no. 682978 - EXCITERS, and from  LASERLAB-EUROPE,
grant agreement no. 284464, EC's Seventh Framework Programme. We acknowledge the financial support of the French National Research Agency through ANR-14-CE32-0014 MISFITS, ANR-14-CE32-0010 XSTASE and IdEx Bordeaux LAPHIA (ANR-10-IDEX-03-02). JC and RT acknowledge financial support from the LABEX PlasaPar project, the program ``Investissements d'avenir'' under the reference ANR-11-IDEX-0004-02, the program ANR-15-CE30-0001-01-CIMBAAD. S.B. acknowledge the NSERC Vanier Scholarship. The data presented in the study are available from the corresponding author on reasonable request. 
 
\section*{Supplementary materials}
Supplementary Text\\
Figs. S1 to S15\\
References \textit{(1-62)}

\clearpage

\begin{center}
    \begin{LARGE}
\textbf{Supplementary Materials}
    \end{LARGE}
\end{center}

\section{Experimental details}

\subsection{Experimental setup}

\begin{figure}[h!]
\begin{center}
\includegraphics{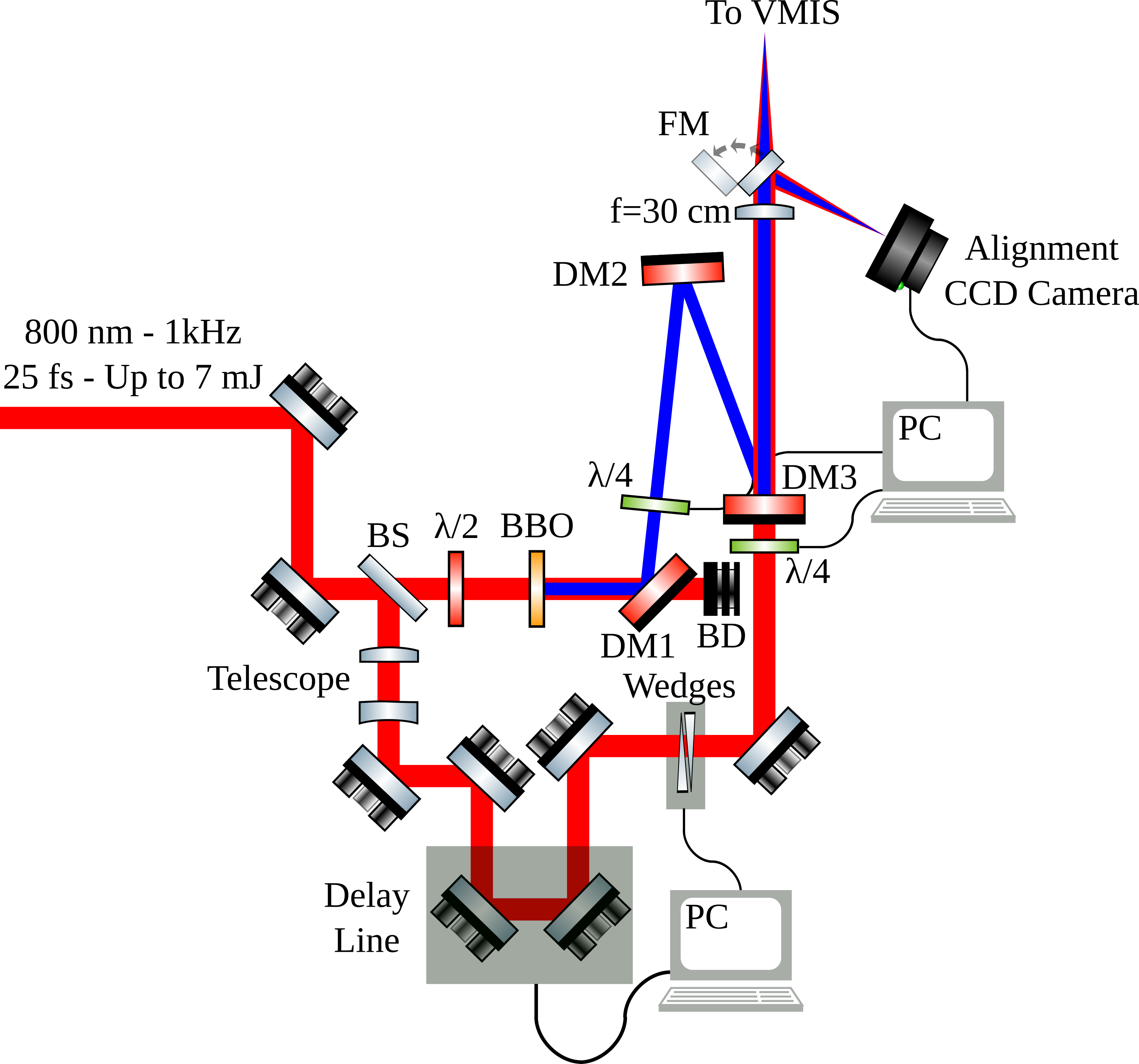}
\caption{Scheme of the optical setup. BS is a 80/20 beamsplitter; $\lambda /2$ and $\lambda /4$ are half- and quarter-wave plates, respectively; BBO is a type-I 200 $\mu$m thick $\beta$-Barium borate crystal; DM are dichroic mirrors; FM is a flip mirror and BD is a beam dump. The setup is not at scale.}
\label{setup}
\end{center}
\end{figure}

A scheme of the experimental setup is shown in figure \ref{setup}. It consists of a standard Mach-Zehnder interferometer. The incoming beam is delivered by the Aurore laser system at CELIA, which provides 800 nm 25 fs pulses at 1 kHz, with up to 7 mJ energy. In one of the two arms, we frequency double the pulses by using a type-I 200 $\mu$m thick BBO crystal. The remaining 800 nm is filtered out by reflection on two dichroic mirrors. In the 800 nm arm, a delay stage is installed to temporally overlap the two beams. The control of the attosecond delay is achieved by a pair of wedges rather than by translating mirrors. This enables us to convert a rather large translation motion of one wedge into small delays (1 micron for 67 attoseconds, while a 1 micron translation of the mirrors would induce a 6.7 femtoseconds delay), releasing some constraints on the translation stage accuracy and repeatability. 

Motorized quarter-wave plates are placed in both arms of the interferometer,  allowing to fully control the polarization of each color independently. After the quarter-wave plates, all reflections are at $\sim 0^\circ$ to avoid polarization state artifacts. The 400 and 800 nm beams are recombined using a dichroic mirror and are focused into the interaction region of the Velocity Map Imaging Spectrometer. To compensate chromatic aberration induced by the lens, we have installed a lens telescope in the 800 nm arm, which allows us to focus both 400 nm and 800 nm at the exact same position in the spectrometer. In all presented experiments, the 400 nm/800 nm delay was scanned over a range of $\sim$ 6700 as by steps of 133 as.  

The enantiopure chiral samples (Sigma Aldrich, 98$\%$ purity for (1R)-(+)-camphor and 95$\%$ purity for (1S)-(-)-camphor) were heated in an oven at 60$^\circ$C, carried to the VMI by a 80$^\circ$C heated line, and conducted under vacuum by a 100$^\circ$C heated tube to a 250 $\mu m$ nozzle located 7 cm away from the interaction zone. No carrier gas was used. The pressure in the interaction chamber was typically 2$\times 10^{-7}$ mbar. The velocity distribution of the photoionized electron was projected onto a set of dual microchannel plates and imaged by a phosphor screen and a CCD camera.

\newpage

\section{Theory}

\subsection{Equivalence between ATI and RABBIT electron interferometry }

The measurement of photoelectron wavepacket spectral amplitudes and phases is usually performed using the RABBIT technique \cite{paul01-1}. In RABBIT, the ionizing radiation is a comb of XUV harmonics. The photoionization is driven by single photon absorption since the photon energy of the harmonics is usually greater than the ionization potential of the target. An additional weak IR pulse is used to produce photoelectron sidebands. The oscillation phase of those sidebands as a function of the XUV-IR attosecond delay provides the information which is needed to reconstruct the temporal dynamics of the photoionization. 

The scheme used in our experiments is similar but the single-photon ionization by high-harmonics is replaced by Above-Threshold Ionization (ATI) by a UV (400 nm) laser field \cite{zipp14}. This choice was motivated by two key elements:  (i) the polarization state of the UV pulse is easily tunable from linear to circular, and (ii) ATI produces much simpler photoelectron spectra in large molecules with many molecular orbitals lying energetically close to each other. In this section of the SM, our goal is to show that our ATI electron interferometry scheme allows us to retrieve equivalent information than in a RABBIT experiment. 

As introduced by Gruson \textit{et al.} \cite{gruson16}, spectrally resolved RABBIT (Rainbow-RABBIT) enables the direct measurement of spectral phase jumps in the vicinity of a resonance. In this case, the \textit{n}-th order harmonic of the XUV comb photoionizes the target close to an autoionizing resonance. The spectral amplitude and phase of the corresponding photoelectron peak are thus strongly modulated by the presence of the resonance. The adjacent harmonics (\textit{(n-1)}-th and \textit{(n+1)}-th orders) photoionize the target in a flat region of the continuum (\textit{i.e.} away from resonances). The spectral amplitude and phase of the associated photoelectron peak are thus unstructured. The absorption/emission of an additional IR photon from both peaks leads to the formation of a sideband. The sideband results from the interference of the resonant photoelectron peak (\textit{n-th}) with the non-resonant adjacent peak. The spectral phase of the sideband thus directly provides the spectral phase of the resonant photoelectron peak, since it is heterodyned by a peak with a flat phase.

Our case is quite similar, but the ATI peaks share some common transition pathways. If a resonance is present at the energy of the (\textit{n-th}) ATI peak, its signature may be carried to the (\textit{(n+1-th)}) ATI peak, which could not be used as a phase reference to retrieve the phase jump. It is thus important to answer the following question: does the effect of the resonance propagate through the ATI peaks, or is it localized on the resonant ATI peak only?  

\begin{figure}[h]
\begin{center}
\includegraphics[scale=1]{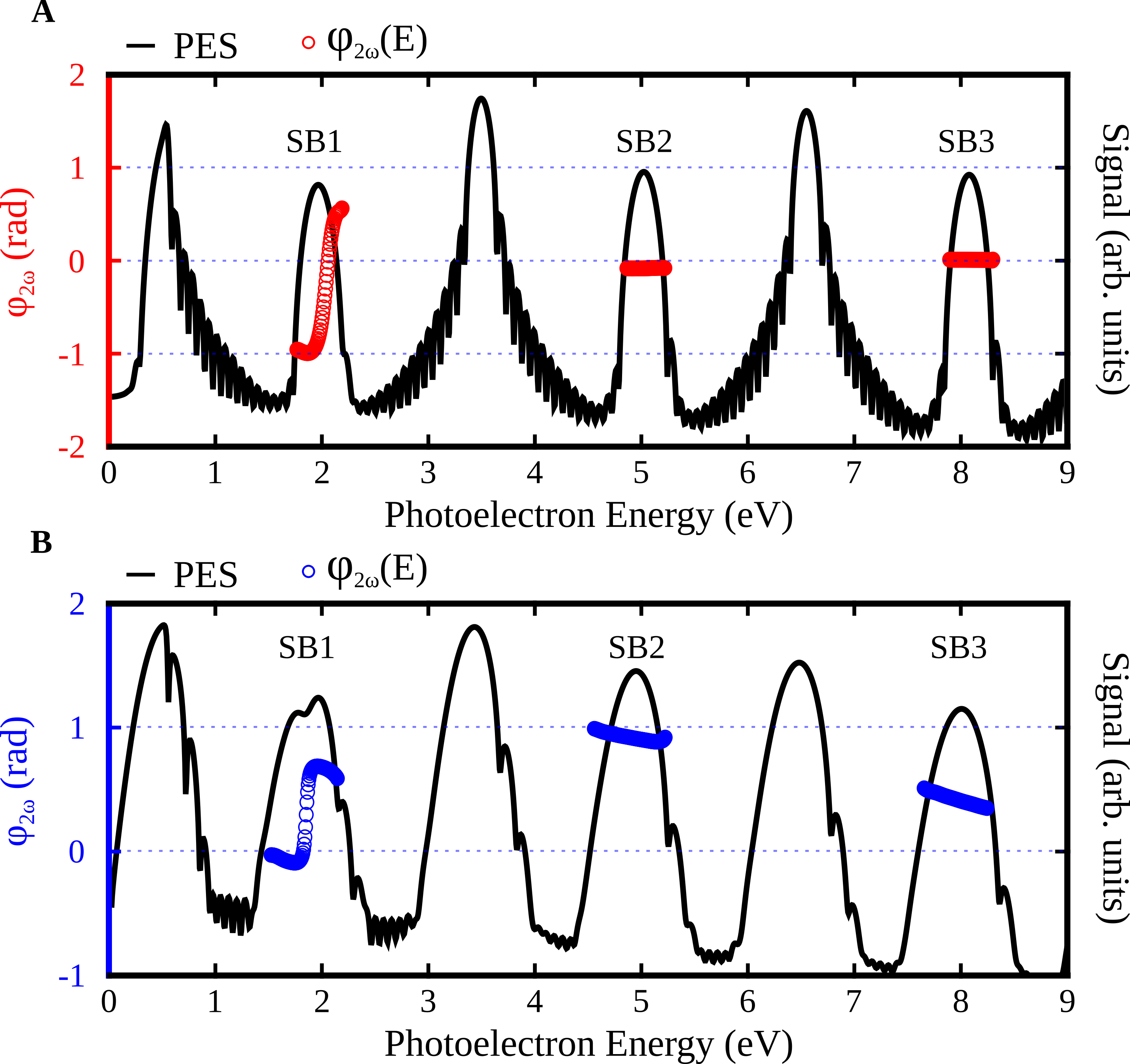}
\caption{Photoelectron interferometry in the RABBIT (\textbf{A}) and ATI (\textbf{B}) regimes. The photoelectron spectra are shown in black. The spectrally-resolved $2\omega$-oscillation phase of the sidebands in the RABBIT regime (A) is shown in red, and in the ATI regime (B) is shown in blue.}
\label{TDSE_LCPMR}
\end{center}
\end{figure}

To do so, we solved the time dependent Schr\"odinger Equation for a 1-D model atom, represented by a short-range Gaussian potential where we added symmetric barriers in order to induce a shape resonance into the continuum, as in \cite{gaillac16}. The parameters of the potential were adjusted to approximately match the ionization potential of the molecule as well as the energy of the resonance encountered in the experiment. The propagation was performed on a spatial grid using the Crank-Nicolson algorithm in the velocity gauge. The photoelectron spectrum was extracted from the final wave function thanks to the window operator.  

We used two kinds of electric field, both composed of harmonics of the fundamental IR field, whose wavelength is set at $806.4$ nm ($\hbar\omega=1.5375$ eV). In the first case, dubbed "ATI", we ionized the system with only the second harmonic ($2\omega$) at an intensity of $I_{2\omega}= 7 \times10^{12}$ W/cm$^{-2}$  and a pulse duration of FWHM $\approx20$ fs in intensity. In the second case, called "RABBIT", the ionizing field was an attosecond pulse train of the same duration, made from harmonics 6, 8, 10, 12 and 14  of the IR frequency $\omega$ in order to reach the same energies as in the "ATI" case. For the sake of simplicity, we considered that all harmonics were synchronized and of the same intensity $I_H= 3 \times10^{8}$ W/cm$^{-2}$.

In order to observe sidebands, we added a weak IR field ($I_{\omega}= 8 \times10^{10}$ W/cm$^{-2}$), with the same duration, and we varied its phase $\delta\equiv\omega\tau$. Following the procedure introduced in \cite{gruson16}, the photoelectron spectra $S(\delta)$ as a function of delay $\delta$ were analysed for each energy, using the following fit:

\begin{equation}\label{fit}
    S(\delta)= \overline{S}\left[1-B_2cos(2\delta+\varphi_2)-B_4cos(4\delta+\varphi_4)\right],
\end{equation}

where $B_2$ and $B_4$ are proportionality coefficients and $\varphi_2$ and $\varphi_4$ are the phases of the $2\omega$- and $4\omega$-oscillations. 

Note that in the "ATI" case, we analyzed the spectra in only one direction, i.e. $x>0$. This is compulsory as the parity of the sideband final state is different, whether the path leading to the sideband comes from the ATI peak below or above. Therefore, there would be no sideband oscillation if the spectra were taken over the all $x$ space. This is not the case in the "RABBIT" case, because the sideband parity is always "even", as invariably reached by a two-photon transition. 

As shown in figure \ref{TDSE_LCPMR}, both RABBIT and ATI techniques show a large spectral phase jump across the bandwidth of the SB1, while it is mostly flat for SB2 and SB3. This results confirm that the effect of the resonance does not propagate above the resonant ATI peak. This observation validates that our experimental interferometric ATI scheme is well suited for the measurement of the spectral amplitude and phase of photoelectron wavepackets. Beyond this similarity, we observe some differences in the evolution of the spectral phases: the phase jumps are slightly different, and a linear phase appears on each higher ATI sidebands as well as between the ATI sidebands. These are the signature of differences between the single photon and multiphoton ionization dynamics. 

\newpage

\subsection{Calculation of differential Wigner delays in one-photon ionization of camphor molecules}

Atomic units will be used throughout this section unless otherwise stated.

\subsubsection{The theoretical approach}

In the framework of the dipolar approximation, ionization probabilities and cross sections depend on the matrix element which, when
defined in the molecular frame, reads 
\begin{equation}
d^{(mol)}_{\bf{k'}}=<\Psi_{\bf{k'}}^{(-)}|{\bf r.E}|\Psi_i>
\label{dipole_MF}
\end{equation}
where ${\bf E}$ is the electric field associated to the incident radiation, $\Psi_i$ is the initial molecular state, and $\Psi_{\bf{k'}}^{(-)}$
is the stationnary ingoing scattering state describing an electron ejected in the direction $\bf{\hat{k'}}$ in the molecular frame. ${\bf r.E}$
is the dipolar interaction term, expressed in the length gauge, where ${\bf r}$ stands for the vectorial coordinate of the active electron. 
In this respect, we work in a single active electron picture where inner electrons remain frozen throughout the interaction and are accordingly 
described by the same molecular orbitals in both $\Psi_i$ and $\Psi_{\bf{k'}}^{(-)}$. 

The optimization of molecular geometry and $\Psi_i$ result from Density Functional Theory calculations using the GAMESS-US package \cite{schmidt93} 
with the 6-311++G** underlying Gaussian basis \cite{krishnan80} and CAM-B3LYP exchange-correlation functional \cite{yanai04}. 

The electric field, circularly polarized, is naturally defined in the laboratory frame so that the dipolar interaction term must be transposed
into the molecular frame as
\begin{eqnarray}
{\bf r.E}=x \pm i y &=& \sqrt{\frac{2\pi}{3}} r (a Y^{-1}_1(\Omega) + b Y^{1}_1(\Omega) ) \nonumber \\
&=&\sqrt{\frac{2\pi}{3}} r \sum_\nu{ (a {\cal D}_{\nu -1}^{(1)} + b {\cal D}_{\nu 1}^{(1)} ) Y^{1}_\nu(\Omega') }
\label{transform_rE}
\end{eqnarray}
where we have assumed $||{\bf E}||=1$ and $Y^m_l$ are usual spherical harmonics defined either in the laboratory frame (with argument $\Omega$) 
or in the molecular frame (with argument $\Omega'$). The coefficients ($a,b$) are (0,-2) in the $x + i y$ case and (2,0) in the $x - i y$ one. 
${\cal D}_{m n'}^{(l)} \equiv {\cal D}_{m n'}^{(l)}(\alpha,\beta,\gamma)$ are Wigner matrix elements \cite{amati60} related to the 
rotation, in terms of the Euler angles $(\alpha,\beta,\gamma)$, that brings the laboratory frame in coincidence with the molecular one. 

The multi-center scattering state $\Psi_{\bf{k'}}^{(-)}$ is expanded in terms of partial waves \cite{dill74}
\begin{equation}
\Psi_{\bf{k'}}^{(-)}({\bf r})=\sum_{l,m} {i^l e^{-i \sigma_l} \Psi_{k'lm}^{(-)}({\bf r}) Y_l^{m*}(\bf{\hat{k'}}) }
\label{partialwaves}
\end{equation}
where $\sigma_l$ is the Coulomb phase shift for electron wavevector $k'$ and angular momentum $l$. The complex $\Psi_{k'lm}^{(-)}({\bf r})$ states, 
which fulfill appropriate boundary conditions, are related to real states $\Psi_{k'lm}({\bf r})$ through the transformation
\begin{equation}
\Psi_{k'lm}^{(-)}({\bf r})=\sum_{l'm'} {({\bf I}+i{\bf K})_{lm,l'm'} \Psi_{k'l'm'}({\bf r}) }
\label{normalizationS}
\end{equation}
where ${\bf I}$ is the unitary matrix and ${\bf K}$ is the so-called ${\bf K}$-matrix (see below). The $\Psi_{k'lm}$ are real solutions of 
the Schr\"odinger equation $H \Psi_{k'lm} = \epsilon \Psi_{k'lm}$, with $\epsilon=k'^2/2$.

To solve the Schrödinger equation, we employ an approximate form of the potential felt by the electron ejected from neutral camphor, in terms of so-called ElectroStatic Potential ESP-charges 
\cite{besler90}, which basically consist of non-integer charges $Z_i^{eff}$ located on the nuclei of the molecule, so that

\begin{equation}
V({\bf r})=-\sum_i {\frac{Z_i}{|{\bf r} - {\bf R}_i|} + \int{\frac{\rho({\bf r'})}{|{\bf r} - {\bf r'}|} }} \sim -\sum_i {\frac{Z_i^{eff}}{|{\bf r} - {\bf R}_i|}}
\label{eqESP}
\end{equation}
on Van der Walls surfaces surrounding the molecule. $Z_i$ and ${\bf R}_i$ are the real charges and locations of the nuclei while $\rho({\bf r})$ is the core electron density which does not include the density associated to the active electron. Therefore $\lim_{r \rightarrow \infty}V({\bf r}) = -1/r$. Subsequently to a single-center decomposition of $V({\bf r})$ onto spherical harmonics \cite{decleva94,abusamha10}, the
coupled-channel Schr\"odinger problem is solved using the renormalized Numerov method of Johnson \cite{johnson78}. The ${\bf K}$-matrix elements are
defined in the asymptotic $r$-region according to  
\begin{equation}
\Psi_{k'lm}({\bf r}) \sim \frac{1}{\sqrt{\pi k'}r} \sum_{l''m''} {(\sin(\theta_{l''})\delta_{l'm',l''m''} +
\cos(\theta_{l''})K_{l'm',l''m''})Y_{l''}^{m''}({\bf \hat{r}}) }
\label{Kmatrix}
\end{equation}  
where $\theta_{l''}=k'r-l''\pi/2-(1/k')\ln(2k'r)+\arg\Gamma[l''+1-i/k']$. 

Once the $\Psi_{k'l'm'}$ states and ${\bf K}$-matrix elements are known, we are able to compute the dipolar amplitudes $a_{klm\nu}$  
\begin{equation}
a_{klm\nu}=<\Psi_{klm}^{(-)} | r Y^\nu_1 |  \Psi_i >
\label{amplitudes}
\end{equation}
using eq. (\ref{normalizationS}). According to eqs. (\ref{dipole_MF})-(\ref{partialwaves}), the dipole in the molecular frame is thus simply 
\begin{equation}
d^{(mol)}_{\bf{k'}}=\sqrt{\frac{2\pi}{3}} \sum_{lm\nu} {(-i)^l e^{i\sigma_l} a_{k'lm\nu} (a {\cal D}_{\nu -1}^{(1)} + b {\cal D}_{\nu 1}^{(1)} ) Y^{m}_l(\bf{\hat{k'}}) }.
\label{dipole_MF_bis}
\end{equation}
Performing the inverse rotation on $Y^{m}_l$ to pass from the molecular frame to the laboratory one, we obtain the dipole in this latter
\begin{equation}
d^{(lab)}_{\bf{k}}=\sqrt{\frac{2\pi}{3}} \sum_{lm\nu\mu} {(-i)^l e^{i\sigma_l} a_{klm\nu} (a {\cal D}_{\nu -1}^{(1)} + b {\cal D}_{\nu 1}^{(1)} ) {\cal D}_{m \mu}^{(l)*} Y^{\mu}_l(\bf{\hat{k}}) }
\label{dipole_LF}
\end{equation}
which can be evaluated for any direction of electron ejection $\bf{\hat{k}}=(\theta,\varphi)$ in the laboratory. 

Introducing the simple notation $d^{(lab)}_{\bf{k}}=|d^{(lab)}_{(k,\theta,\varphi)}|e^{i\varphi(k,\theta,\varphi)}$, the Wigner delay can be computed in the case of an oriented molecule as \cite{wigner55,chacon14,goldberger62,baykusheva17} 
\begin{equation}
\tau_W(k,\theta,\varphi)=\frac{\partial}{\partial E} \varphi(k,\theta,\varphi) \equiv \frac{1}{k}\frac{\partial}{\partial k} \varphi(k,\theta,\varphi).
\label{delay_oriented}
\end{equation}
However this delay is defined for a given orientation $\hat{R}$ while the experiment deals with samples of randomly oriented molecules. We thus have to introduce an orientation-averaged delay
$\bar{\tau}_W(k,\theta,\varphi)$, within which the contribution of a particular orientation $\hat{R}$ is weighted by its contribution to the total electron production 
\begin{equation}
{\cal N}(k,\theta,\varphi)=\int{d\hat{R} |d^{(lab)}_{\bf{k}}(\hat{R})|^2},
\label{calN}
\end{equation}
according to \cite{baykusheva17} 
\begin{equation}
\bar{\tau}_W(k,\theta,\varphi) = \int{d\hat{R} \tau_W(\hat{R};{\bf k}) \frac{|d^{(lab)}_{\bf k}(\hat{R})|^2}{{\cal N}(\bf{k})}}.
\label{delay_av}
\end{equation}
In practice, the integrations on $\hat{R}$, with $d{\hat R}=\frac{1}{8\pi^2}d\alpha \sin(\beta)d\beta d\gamma$, are performed by (Simpson) numerical quadratures with angular spacing
$\Delta \alpha=\Delta\beta= \Delta\gamma$ small enough to ensure convergence of the computed $\bar{\tau}_W$ values. 

The one-photon counterpart to the measured differential Wigner delay is then evaluated by subtracting the $\bar{\tau}_W$ values in the forward and backward directions:
\begin{equation}
\Delta \bar{\tau}_W^{f/b}(k)= \bar{\tau}_W(k,0,\varphi) - \bar{\tau}_W(k,\pi,\varphi), 
\label{deltatauf/b}
\end{equation}
the axis $z$ of quantization being collinear with the direction of propagation of the incident direction. 
The delay difference is also inspected between the left and right directions:
\begin{equation}
\Delta \bar{\tau}_W^{l/r}(k)= \bar{\tau}_W(k,\pi/2,\pi) - \bar{\tau}_W(k,\pi/2,0).
\label{deltatauLR}
\end{equation}
Note that similar quantities can be evaluated for a fixed orientation, $\Delta \tau_W^{f/b}(\hat{R};k)= \tau_W(\hat{R};k,0,\varphi) - \tau_W(\hat{R};k,\pi,\varphi)$ and 
$\tau_W^{l/r}(\hat{R};k)= \tau_W(\hat{R};k,\pi/2,\pi) - \tau_W(\hat{R};k,\pi/2,0)$. 

Finally, our calculations also enable to estimate the PhotoElectron Circular Dichroism (PECD) in the forward/backward and left/right directions as
\begin{eqnarray}
PECD^{f/b}(k)&=& 2\frac{{\cal N}(k,0,\varphi) - {\cal N}(k,\pi,\varphi)}{{\cal N}(k,0,\varphi) + {\cal N}(k,\pi,\varphi)} \label{PECDf/b}\\
PECD^{l/r}(k)&=& 2\frac{{\cal N}(k,\pi/2,\pi) - {\cal N}(k,\pi/2,0)}{{\cal N}(k,\pi/2,\pi) + {\cal N}(k,\pi/2,0)} \label{PECDlr}.\\
\end{eqnarray}

\subsubsection{Results}

All the following results have been obtained using $\Delta \alpha=\Delta\beta= \Delta\gamma =\pi/32$ which guaranteed convergence of all computed observables. 

\begin{figure}
\begin{center}
\includegraphics[width=0.80\textwidth, angle =-90]{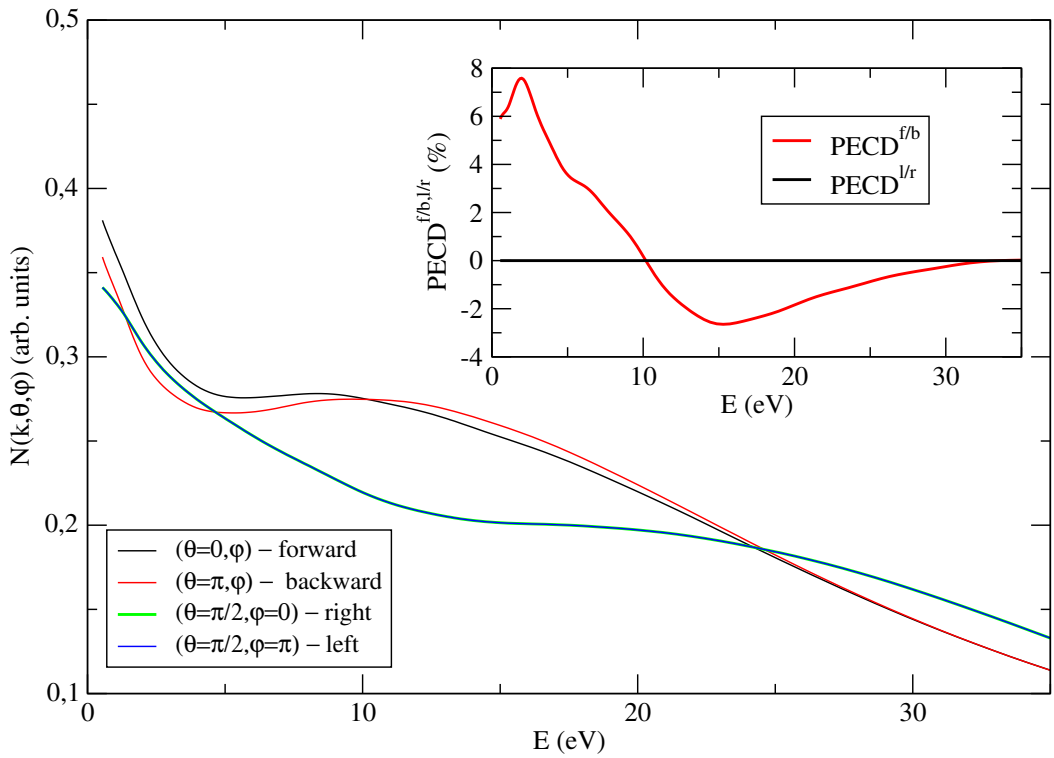}
\caption{Differential electron production ${\cal N}$, defined in eq.(\ref{calN}), along the forward, backward, left and right directions of electron ejection,
for (1S)-(-)-camphor irradiated by a left circularly polarized radiation. In the inset are displayed the PECDs in the forward/backward and
left/right directions.}
\label{fig_si_1}
\end{center}
\end{figure}

We first present in Fig. \ref{fig_si_1} the differential electron production ${\cal N}$, defined in eq. (\ref{calN}), along the forward, backward, left and right directions of electron ejection,
for (1S)-(-)-camphor irradiated by a left circularly polarized radiation. While ${\cal N}$'s are strictly identical in the left and right directions, they significantly differ in the forward and backward directions for electron energies $E$ less than 30 eV. This is a feature commonly observed in photoionization of chiral systems \cite{ritchie76,nahon16}, which gives rise to a sizeable 
$PECD^{f/b}$ while $PECD^{l/r}=0$ for all $E$ (see inset of Fig. \ref{fig_si_1}). Interestingly the $PECD^{f/b}$ changes sign about $E=10$ eV. Since our calculations assume that the nuclei are frozen and consider only the single outermost molecular orbital, such a sign change illustrates kinetic energy effects, i.e. the energy dependence of multiple electron scattering off the chiral potential \cite{ritchie76,nahon16,beaulieu16}. The $PECD^{f/b}$ exhibits a maximum value of $\sim 8\%$ about
$E=2$ eV, and a shoulder shows up in the $E=6-10$ eV region. 

\begin{figure}
\begin{center}
\includegraphics[width=0.80\textwidth,  angle = -90]{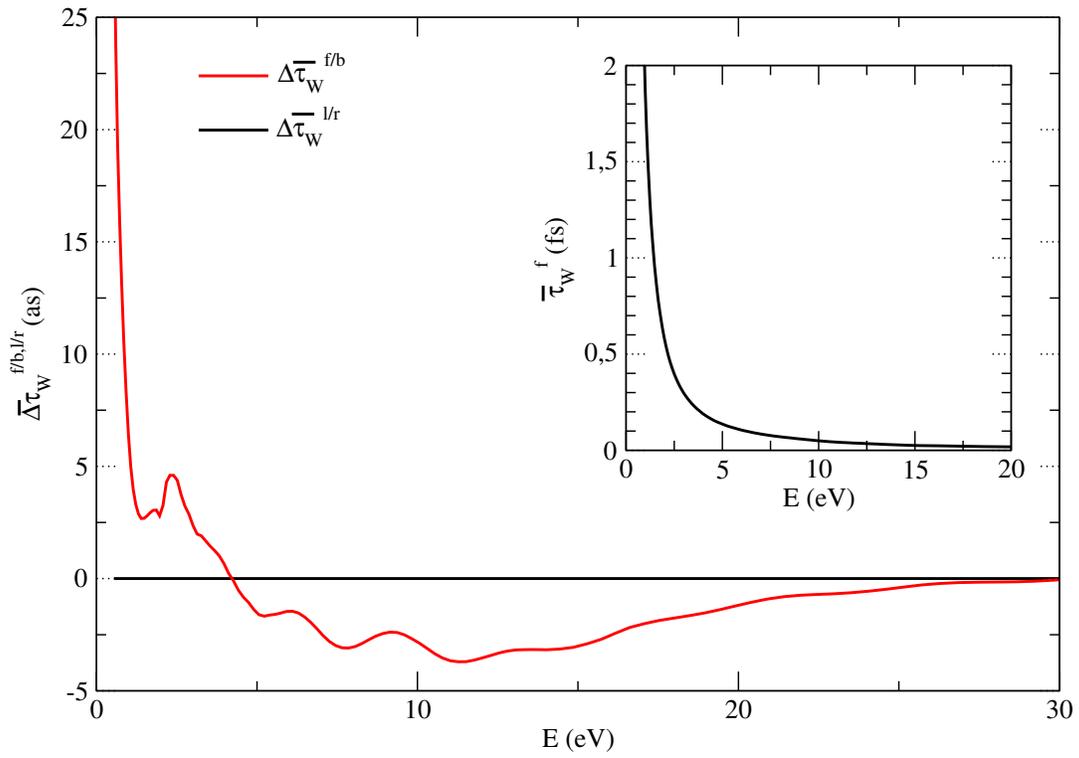}
\caption{Differential Wigner delays, $\Delta \bar{\tau}_W^{f/b}$ and $\Delta \bar{\tau}_W^{l/r}$, defined in eqs. (\ref{deltatauf/b}) and (\ref{deltatauLR}) respectively, for (1S)-(-)-camphor irradiated by a left circularly polarized radiation. The Wigner delay $\bar{\tau}_W^{f}$ 
in the forward direction is illustrated in the inset.}
\label{fig_si_2}
\end{center}
\end{figure}

The differential Wigner delays, $\Delta \bar{\tau}_W^{f/b}$ and $\Delta \bar{\tau}_W^{l/r}$, defined in eqs. (\ref{deltatauf/b}) and (\ref{deltatauLR}) respectively, are
displayed as a function of $E$ in Fig. \ref{fig_si_2}. As for the PECD, we find that $\Delta \bar{\tau}_W^{l/r}=0$ whatever is $E$. By contrast, $\Delta \bar{\tau}_W^{f/b} \ne 0$ and
presents sizeable values in the attosecond range. The forward/backward differential delay is maximum at low $E$, as intuitively expected since low energy electrons spend more time in
the chiral potential. However, it consists in this energy region of a small fraction only of the typical Wigner delay, which is illustrated in the inset of Fig. \ref{fig_si_2}. In practice, 
$\bar{\tau}_W^f \sim \bar{\tau}_W^b \sim \bar{\tau}_W^l \sim \bar{\tau}_W^r$ and the ratio $|\Delta \bar{\tau}_W^{f/b}/\bar{\tau}_W^{f,b}|$ maximizes in the intermediate energy region 
centered about $\sim 10$ eV. Kinetic energy effects also show up in $\Delta \bar{\tau}_W^{f/b}$ through a sign change about $E=4$ eV. But more interestingly, the differential
Wigner delay presents rich spectroscopic features in terms of local maxima/minima for $E > 1$ eV. These features are totally absent in the electron production signals of
Fig. \ref{fig_si_1} and almost invisible in the usual PECD (inset of Fig. \ref{fig_si_1}). In fact, differentiating the PECD with respect to $E$ partially allows retrieving the structures strongly marked in $\Delta \bar{\tau}_W^{f/b}$, but with very small amplitudes. This procedure would thus be difficult on experimental data. By contrast, our present experimental investigation shows that measuring differential delays is now feasible, and will surely be addressed by highly sensitive setups in the next future. In this respect, it seems that the differential delay $\Delta \bar{\tau}_W^{f/b}$ is a very valuable chiral observable which encodes subtle features of the underlying chiral potential, beyond usual (PECD) signatures. 

\begin{figure}
\begin{center}
\includegraphics[width=0.80\textwidth]{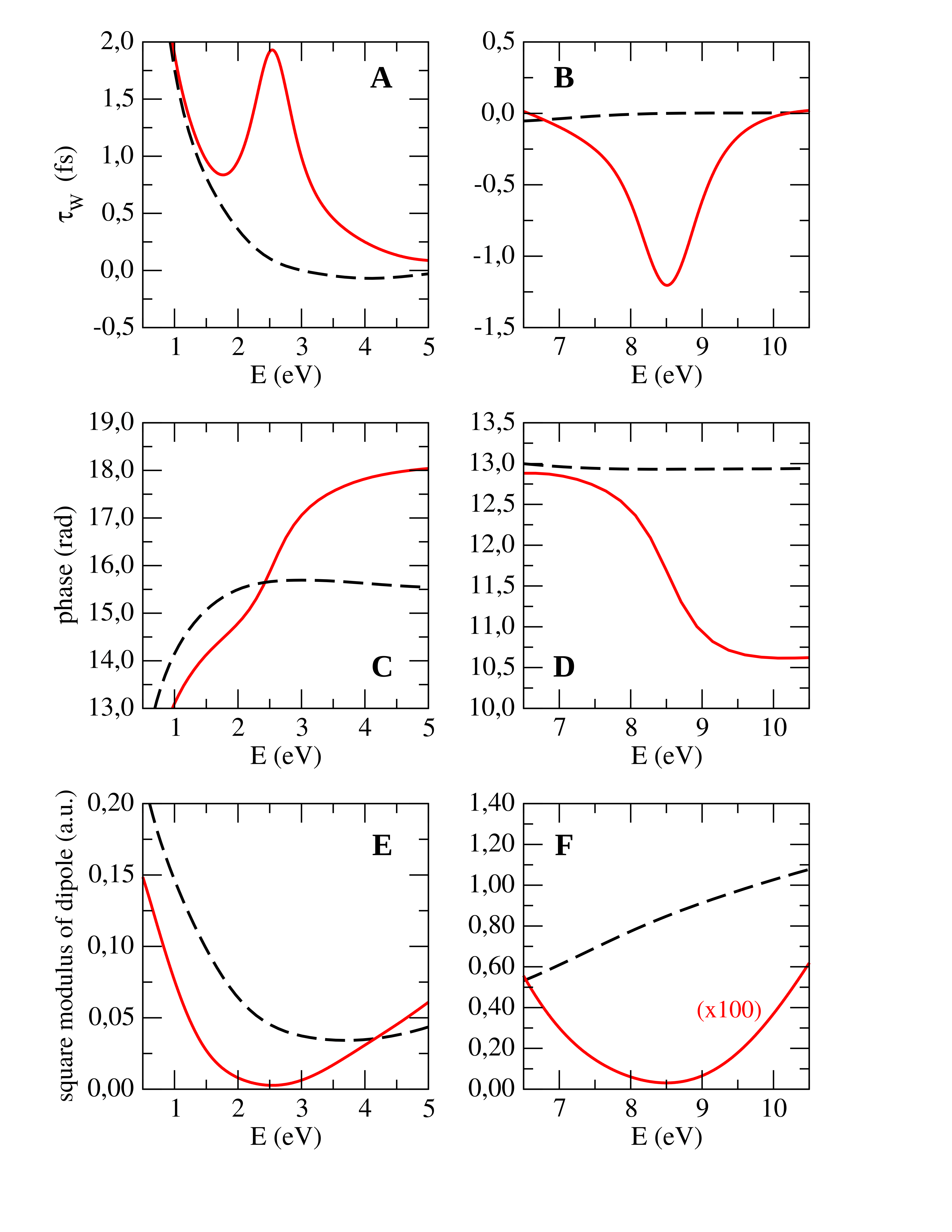}
\caption{(\textbf{A}) Wigner delays, $\tau_W$, in the forward and backward directions for ($\alpha=22.5^\circ,\beta=45^\circ,\gamma=0^\circ$); the (red) continuous line refers to the forward direction and the (black) dashed one to the backward direction.
Associated variations of the phase (\textbf{B}) and square modulus (\textbf{C}) of the dipoles. (\textbf{D-F}) graphs are similar to (A-C), but for
($\alpha=180^\circ,\beta=157.5^\circ,\gamma=0^\circ$).}
\label{fig_si_3}
\end{center}
\end{figure}

We now investigate the origin of the features observed in the differential Wigner delays. $\Delta \bar{\tau}_W^{f/b}$ involves delays averaged on the molecular 
orientations. We thus looked at the delays for all underlying orientations, $\Delta \tau_W^{f/b}(\hat{R};E)$, and observed that some of them have very important values. 
For instance, focusing on the local maximum of $\Delta \bar{\tau}_W^{f/b}$ centered about $E=2.5$ eV in Fig. \ref{fig_si_1}, we found that that the orientation defined by ($\alpha=22.5^\circ,\beta=45^\circ,\gamma=0^\circ$) yields $\Delta \tau_W^{f/b}(\hat{R};E) \sim 2$ fs at 
this energy (see Fig. \ref{fig_si_3}(A)). This important delay difference stems from a phase jump of $\sim\pi$ in the forward direction, in a narrow $E$-range 
where the phase in the backward direction behaves smoothly and thus leads to a vanishing Wigner delay (see Fig. \ref{fig_si_3}(B)). 
Simultaneously the dipole modulus has $\sim 0$ amplitude in the forward direction (see Fig. \ref{fig_si_3}(E)). In other words, 
the important delay in the forward direction is nothing else than the signature of a Cooper minimum \cite{cooper62,cloux15} induced
by the potential shape in the ($\theta=0,\varphi$)-electron direction for ($\alpha=22.5^\circ,\beta=45^\circ,\gamma=0^\circ$) molecular
orientation. This large delay difference, in the fs range, survives orientation averaging, and leads to the maximum of
$\Delta \bar{\tau}_W^{f/b}$ in the as range at $E \sim 2.5$ eV (see Fig. \ref{fig_si_3}).

In our calculations, which are based on the single-active electron approximation and employ a simplified form of the molecular potential
in terms of ESP-charges, all local variations of $\Delta \bar{\tau}_W^{f/b}$ are due to such differential Cooper minima. 
This is illustrated in Figs. \ref{fig_si_3}(D-F) for the oscillation appearing in Fig. \ref{fig_si_2} about $E=8.5$ eV. In this case 
the important negative delay computed in the forward direction for ($\alpha=180^\circ,\beta=157.5^\circ,\gamma=0^\circ$) contributes
positively to the averaged $\Delta \bar{\tau}_W^{f/b}$ because $\sin(\beta)<0$. It is important to note that in the experiment, the multiphoton excitation/ionization preferentially selects some molecular orientations within the sample of ramdomly oriented molecules. Our calculations suggest that this photoselection can have a huge impact on the measured quantities. In the future, it would thus be of great interest to study the differential Wigner time delays within the molecular-frame of chiral molecules.

It has to be noted that not only differential Cooper minima but also differential resonances are amenable to important local variations of $\Delta \tau_W^{f/b}(\hat{R};E)$ because of underlying $\sim \pi$ phase jumps. However, our calculations are based on the single-active electron approximation, which inhibits the occurrence of autoionizing resonances. Shape resonances do not
show up either, at least using our simplified description of the ionic potential in terms of ESP-charges. Nevertheless, it is clear
that fine differential features of chiral molecular potentials can, in general, be probed by measurements and computations of
$\Delta \bar{\tau}_W^{f/b}$. 

\clearpage

\section{Data analysis and error bars}

\subsection{Raw VMI images}

For each time delay, each enantiomer and each polarization state, we record 2D projections of the 3D-PAD, averaged over $\mathrm{\sim 5x10^4}$ laser shots. As mentioned in the main text, these photoelectron distributions are up-down asymmetric. We thus separate the up- and the down- part of the signals, and symmetrize them in order to be able to perform the pBasex inversion. A typical raw VMI image (before any symmetrization or anti-symmetrization) in linear scale (Fig. \ref{raw}(A)) and in logarithmic scale (Fig. \ref{raw}(B)) are presented below. For this example, the UV was linearly polarized, the IR was circularly polarized and the target molecules were (1S)-(-)-camphor. 
\begin{figure}[h!]
\begin{center}
\includegraphics[width=0.80\textwidth]{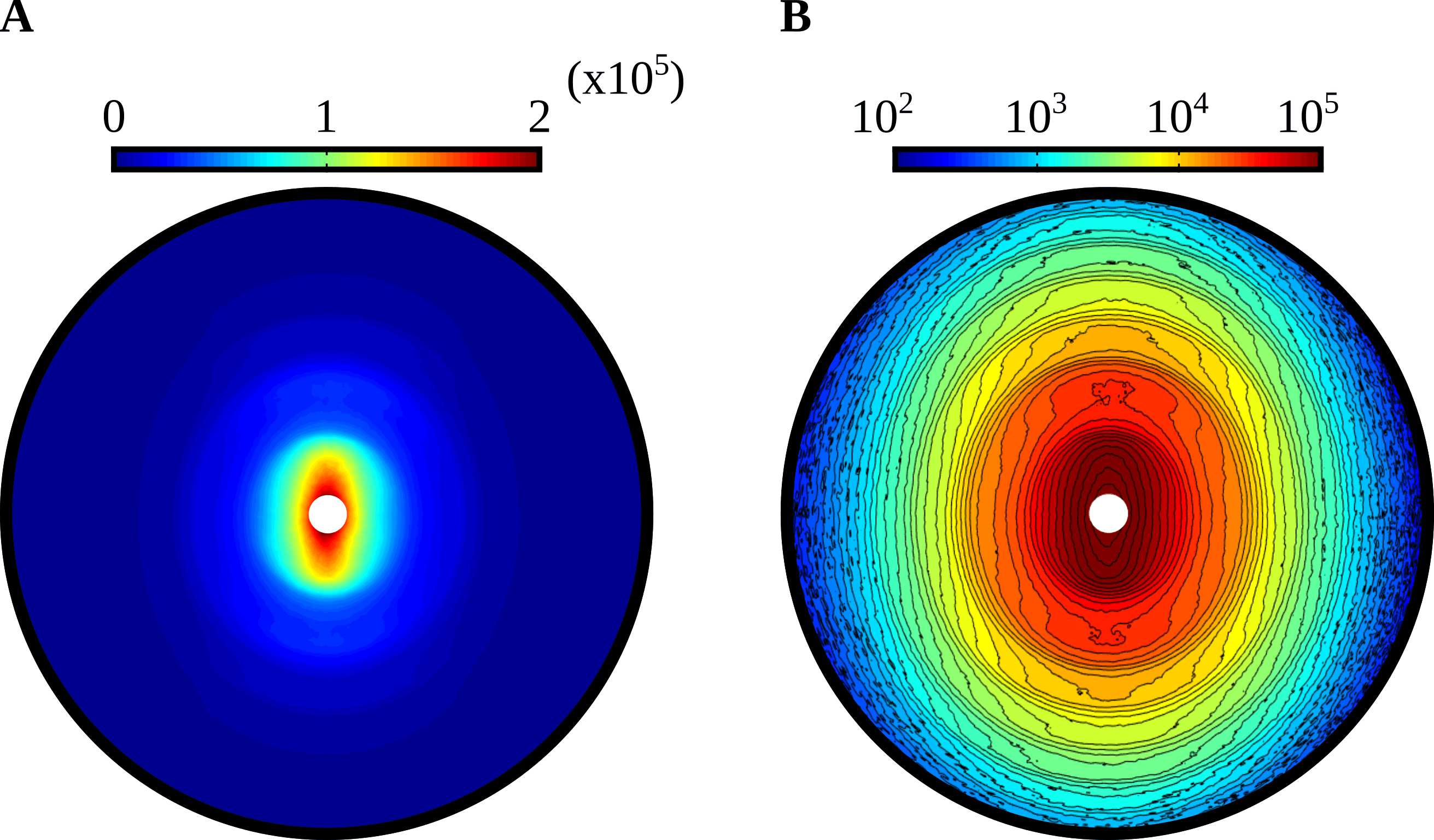}
\caption{Raw experimental VMI images. (\textbf{A}) In linear scale and (\textbf{B}) in logarithmic scale. These images have been recorded using circularly polarized UV and linearly polarized IR fields, in 1S-(-)-camphor. }
\label{raw}
\end{center}
\end{figure}

\subsection{Resonant photoionization of chiral molecules}

This subsection describes the procedure for extracting the error bars of Fig. 4 (B) and (D) of the main paper. 

\begin{figure}[h!]
\begin{center}
\includegraphics[width=0.80\textwidth]{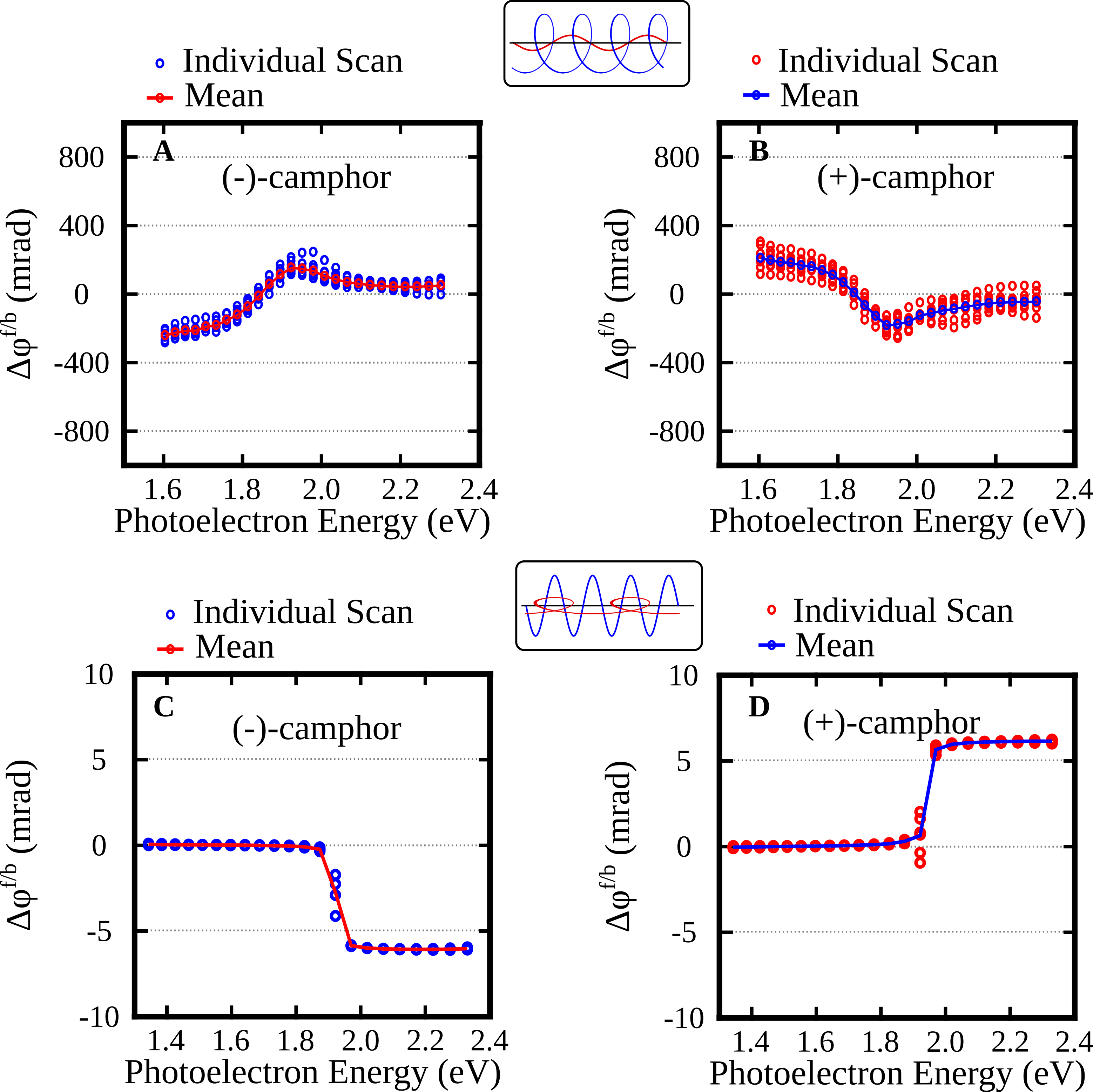}
\caption{Forward/Backward asymmetry of the spectral phase ($\Delta \varphi^{f/b}$) across the resonant sideband. In (\textbf{A}) ((1S)-(-)-camphor) and (\textbf{B}) ((1R)-(+)-camphor), the 400 nm is circularly polarized and the 800 nm is linearly polarized. In (\textbf{C}) ((1S)-(-)-camphor) and (\textbf{D}) ((1R)-(+)-camphor), the 400 nm is linearly polarized and the 800 nm is circularly polarized. For the (1S)-(-)-camphor panels (A,C), the blue dots represent the $\Delta \varphi^{f/b}$ extracted for different scans and the red lines/dots are the mean values. For the (1R)-(+)-camphor panels (B,D), the red dots represent the $\Delta \varphi^{f/b}$ extracted form different scans and the blue lines/dots are the mean values.}
\label{femto}
\end{center}
\end{figure}

For each polarization configuration and for each enantiomer, at least five consecutive delay scans are performed for left and right polarizations. The photoelectron images obtained using left (L) and right (R) polarizations are used to produce a sum image (L+R) and a difference image (L-R). Each photoelectron image is averaged over $\sim 5\cdot10^4$ laser shots. For a non-chiral sample, or non-chiral ionizing light, (L-R) is expected to be zero, while for a chiral sample photoionized with a chiral light it is expected to be non-zero and antisymmetric with respect to the laser propagation axis. On the contrary, L+R images are symmetric with respect to the laser propagation axis. In order to eliminate artifacts due to the imperfect nature of the waveplates and detector, we antisymmetrize the (L-R) images and symmetrize the (L+R) images. 

The (L-R) and (L+R) images are decomposed as a sum of Legendre polynomials \cite{garcia04}. When the laser-molecule interaction is cylindrically symmetric with respect to the detection plane (for instance when both the pump and probe pulse are linearly or circularly polarized), this decomposition procedure enables reconstructing the 3D photoelectron angular distribution from its 2D projection onto the VMI detector. In the present experiment, a combination of circular and linear polarization is used and the cylindrical symmetry condition required to rigorously retrieve the 3D distribution from its projection is lost. This can lead to deviations between the Legendre decomposition and the 3D distribution, in particular in terms of energy smearing. However, in our measurements we detect very sharp structures in the Legendre decomposition: the phase jump across SB1 extends over about 0.1 eV. This indicates that there is no major blurring of the reconstructed distribution, and that it must not deviate much from the 3D distribution. The full 3D distribution could be retrieved by repeating the experiment for different directions of the linearly polarized field and performing a tomographic reconstruction \cite{wollenhaupt09}. This procedure would be prohibitively long with a 1kHz laser system, but will be achievable in a reasonable time using fiber-based lasers at few 100 kHz rate.

We analyze the oscillations of the photoelectron image associated with left polarization (L) by summing the (L-R) and (L+R) decompositions. The R image, obtained by the difference between (L+R) and (L-R), would give perfectly antisymmetric results with respect to the light propagation axis. 

In order to study the statistical dispersion of our data, the phase of the sidebands oscillations is measured for each delay scan, in each quadrant. The forward and backward phases from a given (upper or lower) hemisphere are subtracted to calculate $\Delta \varphi^{f/b}$. Two values are thus obtained for each scan, giving us a total of at least 10 sets of measurements. The $\Delta \varphi^{f/b}$ extracted for each independent measurement as well as their average values are shown in Fig \ref{femto}.

Using Student's statistics, we calculate the 95 $\%$ confidence intervals based on the $n$ individual measurement of $\Delta \varphi^{f/b}$. The confidence intervals are defined as: 

\begin{equation} 
\bigg[\overline{\Delta \varphi^{f/b}} - t \sqrt{\frac{S}{n}} , \overline{\Delta \varphi^{f/b}} + t \sqrt{\frac{S}{n}} \bigg]
\end{equation}

where $\overline{\Delta \varphi^{f/b}}$ is the mean value of all individually measured $\Delta \varphi^{f/b}_i$, $t$ is called the quantile and is a quantity that depends on the confidence interval and the number of individual measurements ($n$), and $S$ is the unbiased estimator of variance, defined as :  

\begin{equation} 
S = \frac{1}{n-1} \sum_{i=1}^{n} \big( \Delta\varphi^{f/b}_i - \overline{\Delta \varphi^{f/b}} \big)^2
\end{equation}

The mean values of the measured f/b asymmetry of the spectral phase, $\overline{\Delta \varphi^{f/b}}$, as well as the 95 $\%$ confidence intervals are presented in figure 4 (B) and (D) of the main paper. We can conclude, based on statistical variability, that our data are accurate and statistically meaningful.

\clearpage

\subsection{Fourier analysis of the sidebands oscillation}

\begin{figure}[h!]
\begin{center}
\includegraphics[width=0.60\textwidth]{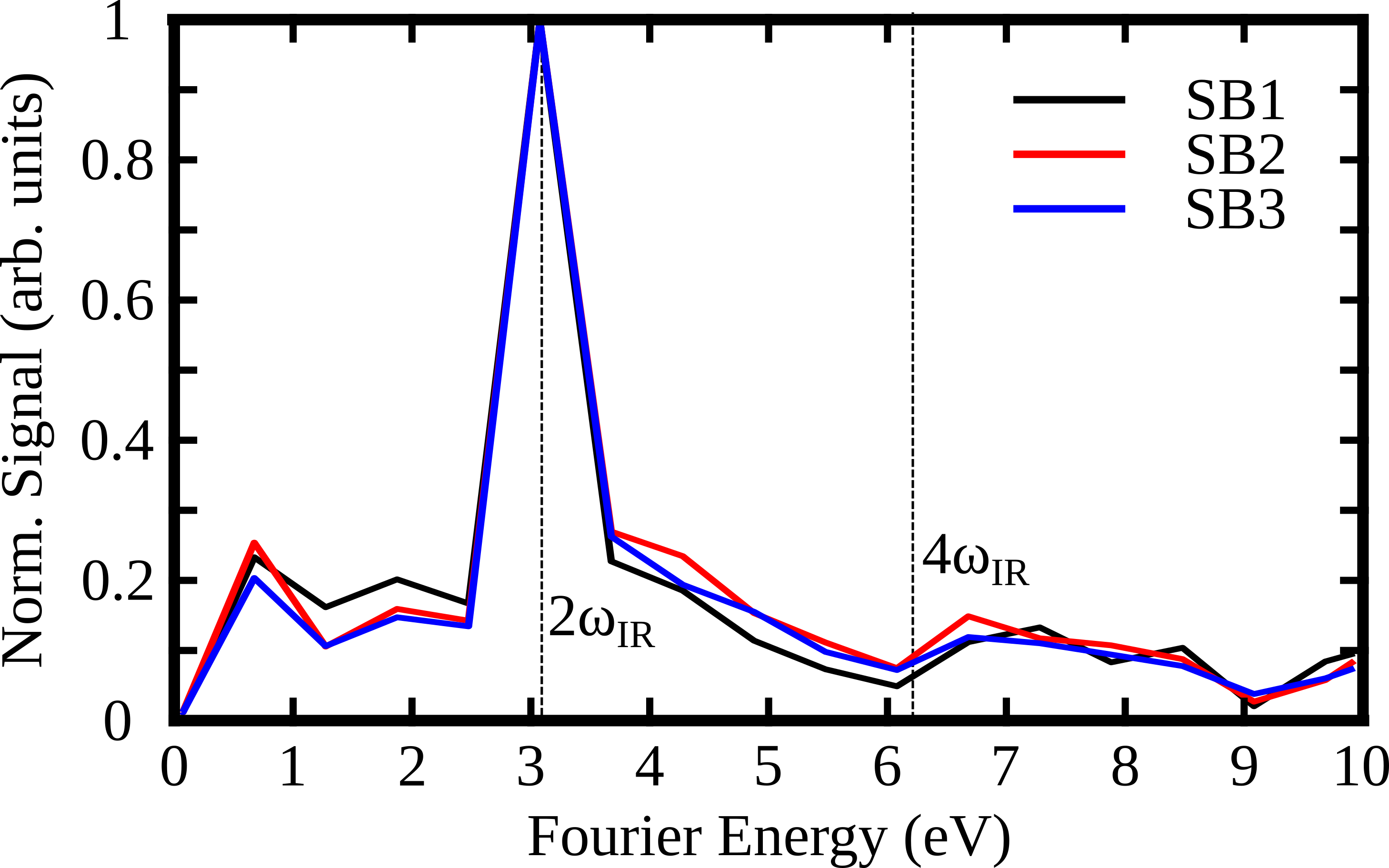}
\caption{Fourier analysis of the temporal oscillation of the SB1 (black), SB2 (red) and SB3 (blue). The photoelectron signal have been integrated over the spectral bandwidth of each SB prior to the Fourier analysis.}
\label{fft}
\end{center}
\end{figure}

The sideband signal as a function of delay is Fourier transformed and shows a single peak at an energy corresponding to a $2\omega_{IR}$ oscillation frequency (Fig. \ref{fft}). This ensures that we are in a RABBIT-like regime and that no higher-order phenomenon affects our phase measurement. Indeed, because ATI peaks also oscillate due to multiple quantum path interferences (\textit{e.g.} for ATI1 interference between [2+1] (3x400nm) and [2+2'] (2x400nm + 2x800nm) paths), one could wonder if this amplitude modulation would affect the phase of the sideband oscillation. A detailed investigation of RABBIT performed by Antoine Camper \textit{et al.} \cite{camper13} demonstrated that when the Fourier transform of the sideband oscillations only showed a single  $2\omega_{IR}$ peak (no $4\omega_{IR}$ or higher orders), the oscillation phase of the sideband was solely determined by the phase difference between adjacent peaks. 

\clearpage

\subsection{Angular-resolved extraction of spectral phase across SB1}

\begin{figure}[h!]
\begin{center}
\includegraphics[width=0.80\textwidth]{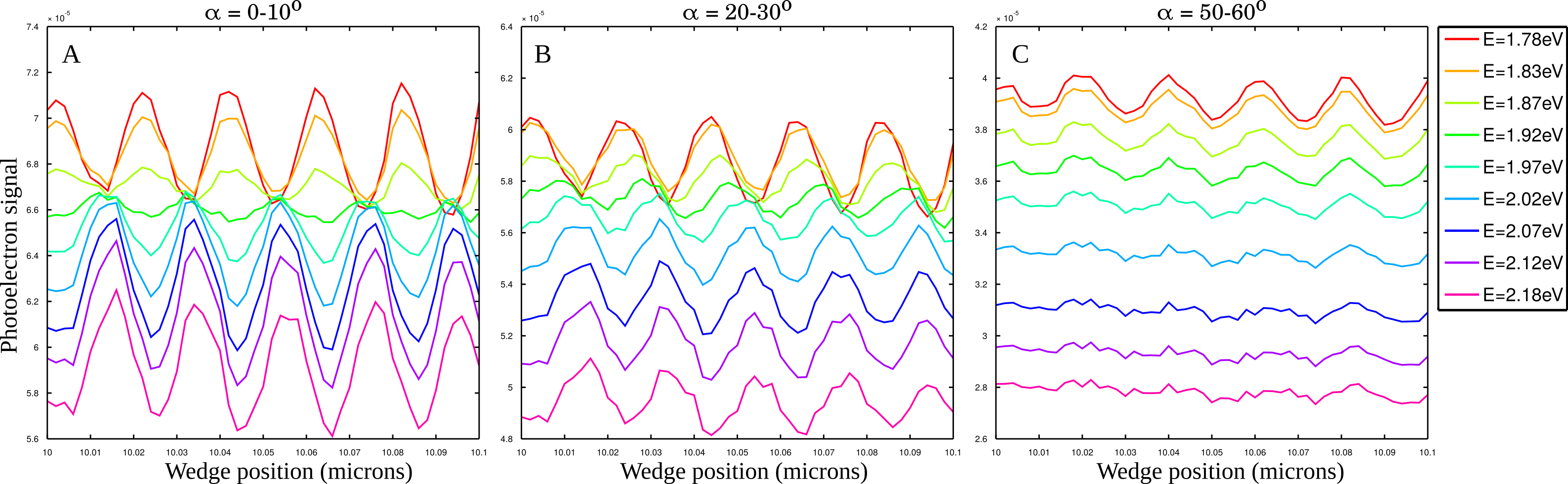}
\caption{Raw oscillations of the SB1 signal as a function of photoelectron energy and ejection angle. The different colored lines represent different photoelectron energy. The electrons are ejected between 0-10$^\circ$ in (\textbf{A}), between 20-30$^\circ$ in (\textbf{B}) and between 50-60$^\circ$ in (\textbf{C}). These data were taken with linearly polarized UV and circularly polarized IR fields, in 1S-(-)-camphor.}
\label{Raw_Oscill}
\end{center}
\end{figure}

In this subsection, we present the oscillation of the SB1 signal as a function of photoelectron energy and ejection angle. In Fig. \ref{Raw_Oscill} (A), we clearly see the steep $\sim \pi$ phase-jump between 1.87 eV and 1.92 eV for electron ejected between 0-10$^\circ$. The magnitude of the phase jump across the resonance is still $\sim \pi$, but is much smoother for electron ejected between 20-30$^\circ$. The phase jump disappear for ejection angle between 50-60$^\circ$. One can notice that the signal-to-noise ratio is very good for electron ejected closer to the laser polarization axis, where the photoelectron signal is maximum. The behavior of the spectral phase as a function of photoelectron ejection angle results in a complex spatio-temporal shaping of the photoelectron wavepacket (see Fig. 4 of the main paper). 

\clearpage

\subsection{Enantiomeric mirroiring in the reconstruction of resonant photoelectron wavepackets}

In this subsection, we present the reconstruction of the autoionizing photoelectron wavepackets for both polarization configurations as well as for both camphor enantiomers. We present the wavepackets along the same ejection angles as shown in the Fig.4 of the main paper. Our aim is to show that even very subtle effects are mirroired when switching the enantiomer. 

\begin{figure}[h]
\begin{center}
\includegraphics[width=0.80\textwidth]{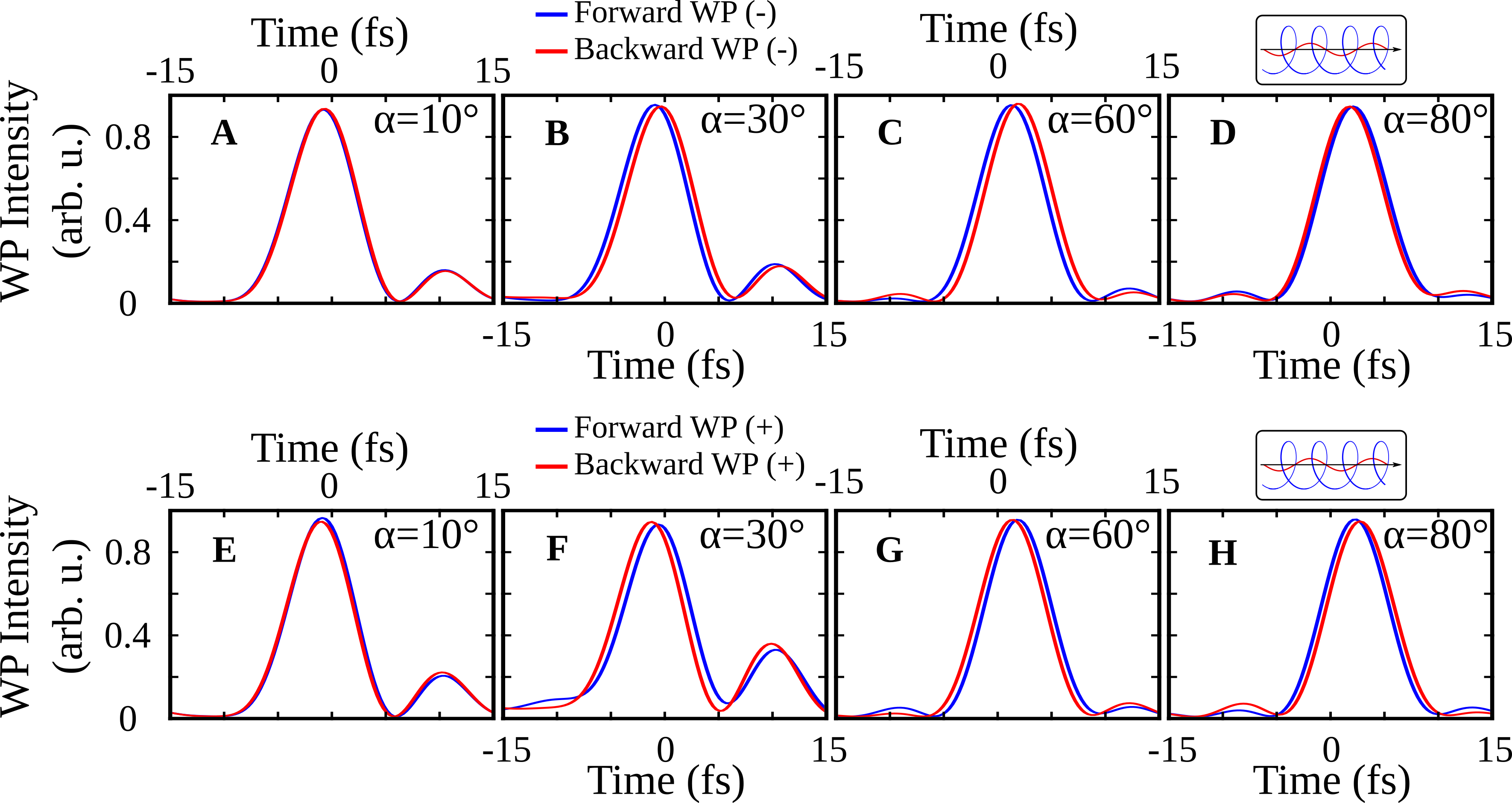}
\caption{Temporal profile of the resonant electron wavepackets emitted in different directions in (1S)-(-)-camphor (top, \textbf{A-D}) and (1R)-(+)-camphor (bottom, \textbf{E-H}), using a circularly polarized UV and a linearly polarized IR field. }
\label{WP_enan1}
\end{center}
\end{figure}

\begin{figure}[h]
\begin{center}
\includegraphics[width=0.80\textwidth]{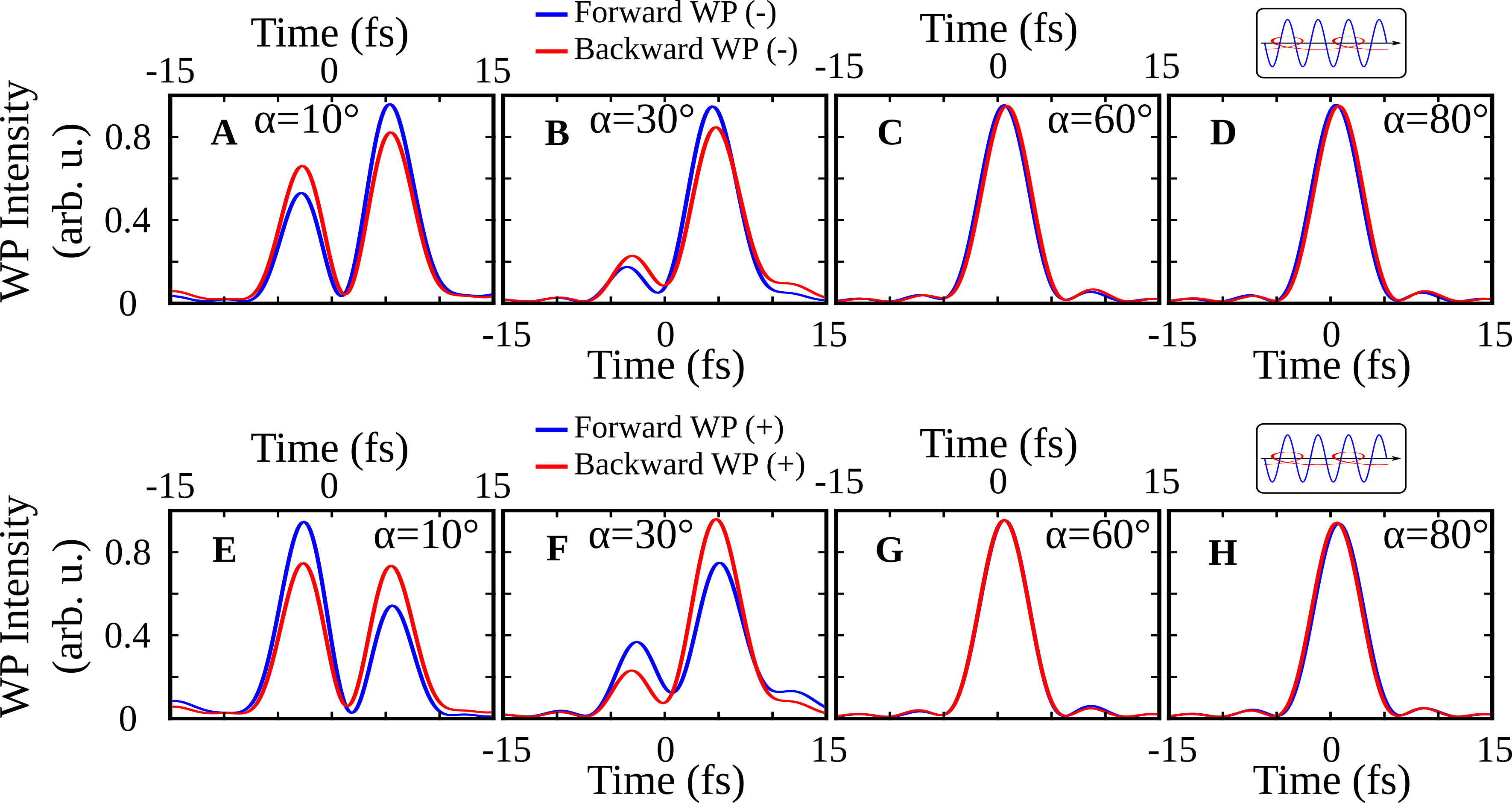}
\caption{Temporal profile of the resonant electron wavepackets emitted in different directions in (1S)-(-)-camphor (top, \textbf{A-D}) and (1R)-(+)-camphor (bottom, \textbf{E-H}), using a circularly polarized UV and a linearly polarized IR field. }
\label{WP_enan2}
\end{center}
\end{figure}

For example, in figure \ref{WP_enan1}, the very small delays between forward and backward wavepacket emitted along $\alpha=30^\circ$ (Fig. \ref{WP_enan1} (B,F)), $\alpha=60^\circ$ (Fig. \ref{WP_enan1} (C,G)), $\alpha=80^\circ$ (Fig. \ref{WP_enan1} (D,H) show a striking mirroiring when switching between enantiomers. It demonstrate the validity of these measurements. A nice enantiomeric mirroiring can also be observed for the case where the UV is linearly polarized and the IR circularly polarized (Fig. \ref{WP_enan2}).

\subsection{Attosecond delays in non-resonant photoionization of chiral molecules}

\begin{figure}[h]
\begin{center}
\includegraphics[width=0.8\textwidth]{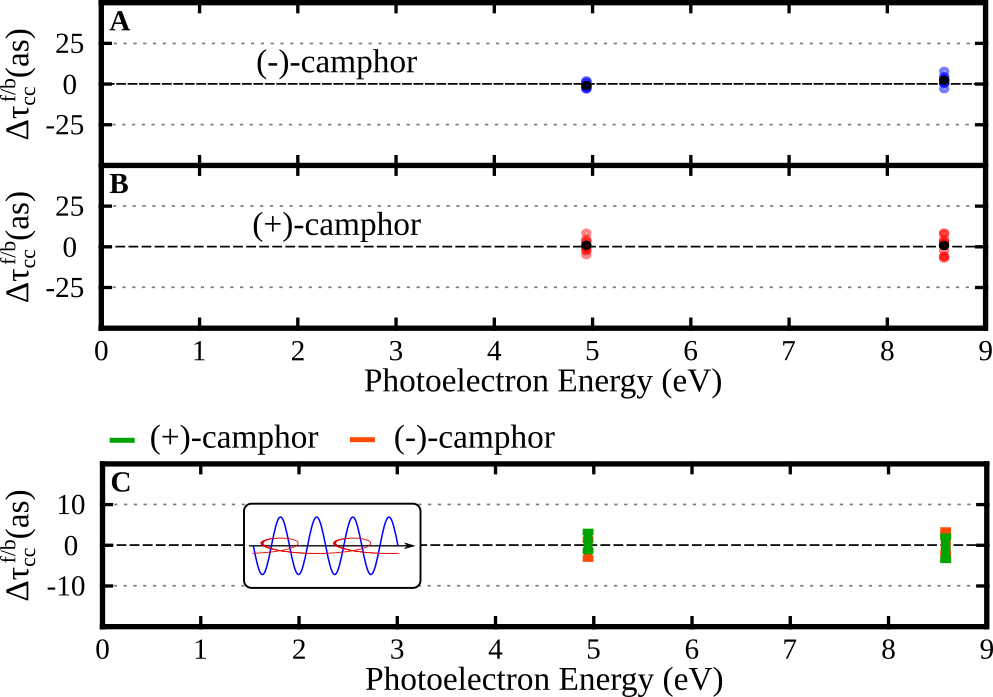}
\caption{Forward/Backward asymmetry in the attosecond photoionization delay ($\Delta \tau^{f/b}$) for the two non-resonant sidebands, when using linearly polarized UV and circularly polarized IR. In (\textbf{A})/(\textbf{B}) we present the $\Delta \tau^{f/b}$ extracted from each independent scans that we have performed in (1S)-(-)-camphor/(1R)-(+)-camphor, respectively.  In (A), the blue dots represent the $\Delta \tau^{f/b}$ extracted from different scans and the black dots are the mean value. In (B), the red dots represent the $\Delta \tau^{f/b}$ extracted from different scans and the black dots are the mean value. In (\textbf{C}), we present the enantiomeric $\Delta \tau^{f/b}$, as defined in equation \ref{enantiomeric}.}
\label{atto2}
\end{center}
\end{figure}

We will describe the procedure to extract the f/b asymmetry of the attosecond delays in non-resonant photoionization, \textit{i.e.} for SB2 and SB3. To do so, we have used the same data sets as in the last section. For each measurement, we calculated the $\Delta \tau^{f/b}$, by using the subtracted forward and backward $2\omega_{IR}$ oscillation phases, which has been extracted from signals integrated over the sideband bandwidth, as in the conventional RABBIT technique. The individual $\Delta \tau^{f/b}$ for SB2 and SB3 are shown as blue dots in figure \ref{atto2} (A) for (1S)-(-)-camphor and as red dots in figure \ref{atto2} (B) for (1R)-(+)-camphor. In both cases, the black dots are the mean values. The value lies close to zero for both enantiomers and both sidebands. When changing the enantiomer, the results should be perfectly mirrored (opposite). Any deviation from perfect mirroring could be ascribed to statistical errors (different S/N ratio for each enantiomer), to systematic error (detector inhomogeneity or polarization state artifact, for example) or to different enantiopurity for each sample. To account for those unwanted detrimental effects, and to estimate more precisely the $\Delta \tau^{f/b}$, we redefine  $\Delta \tau^{f/b}$ as: 

\begin{equation} 
\Delta \tau^{f/b} = \frac{1}{2}\big( \Delta \tau_{(+)}^{f/b} - \Delta \tau_{(-)}^{f/b} \big) 
\label{enantiomeric}
\end{equation}
  
where $\Delta \tau_{(+)}^{f/b}$ and $\Delta \tau_{(-)}^{f/b}$ are the experimentally measured $\Delta \tau^{f/b}$ for (1R)-(+)-camphor and (1S)-(-)-camphor, respectively. Note that after this procedure, we have removed all the unwanted experimental induced errors and the $\Delta \tau^{f/b}$ are perfectly opposite for (1R)-(+)- and (1S)-(-)-camphor. After doing this treatment, we calculate the 95 $\%$ confidence interval error bars using the Student's statistics procedure described in the previous section. The  $\Delta \tau^{f/b}$ for (1R)-(+)-camphor (green dots) and (1S)-(-)-camphor (orange dots) are presented on figure \ref{atto2} (C). Note that for this case, the UV is linearly polarized and the IR is circularly polarized.

One can see that for both non-resonant SB2 and SB3, the $\Delta \tau^{f/b}$ is found to be zero. Again, only the weak IR field is circularly polarized, so the f/b asymmetry is solely broken during 'measurement' step. Our data suggest that for non-resonant photoionization, the 'measurement' step induces the same delay in the forward than in the backward direction, with respect to the light propagation direction. 

\begin{figure}[h]
\begin{center}
\includegraphics[width=0.8\textwidth]{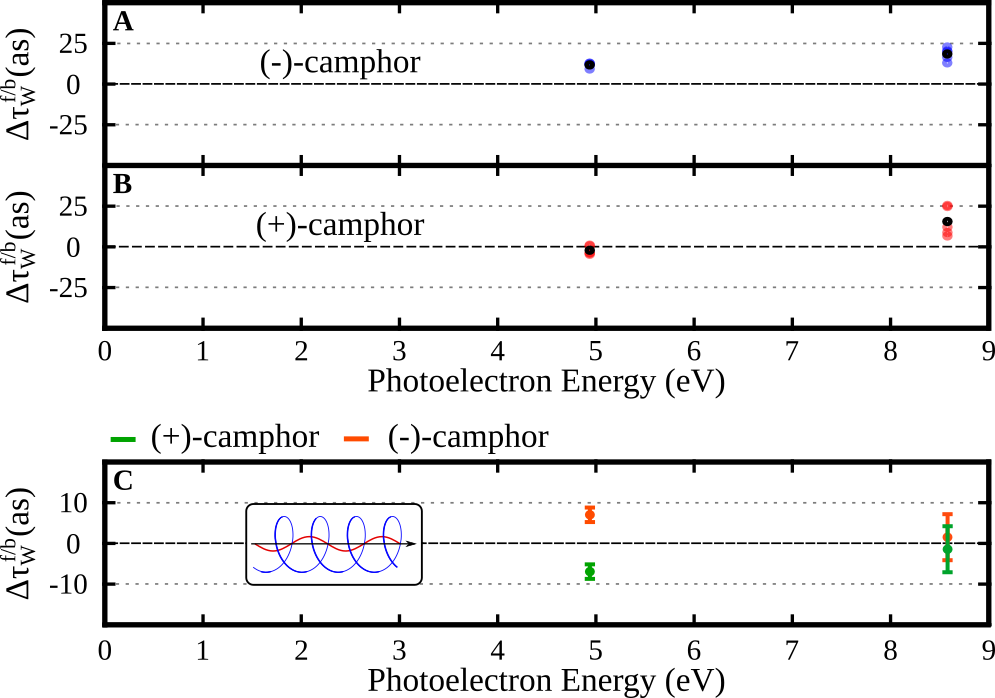}
\caption{Forward/Backward asymmetry of the attosecond photoionization delay ($\Delta \tau^{f/b}$) for the two non-resonant sidebands, when using circularly polarized UV and linearly polarized IR. In (\textbf{A})/(\textbf{B}) we present the $\Delta \tau^{f/b}$ extracted from each independent scans that we have performed in (1S)-(-)-camphor/(1R)-(+)-camphor, respectively.  In (A), the blue dots represent the $\Delta \tau^{f/b}$ extracted from different scans and the black dots are the mean value. In (B), the red dots represent the $\Delta \tau^{f/b}$ extracted from different scans and the black dots are the mean values. In (\textbf{C}), we present the enantiomeric $\Delta \tau^{f/b}$, as defined in equation \ref{enantiomeric}.}
\label{atto1}
\end{center}
\end{figure}

We now turn our attention to the case when the UV is circularly polarized and the IR is linearly polarized. In that case, because the IR is linear, it cannot induce any f/b asymmetry in the 'measurement' step, which means that the measured delays will only reflect the f/b asymmetry of the Wigner time delay. Our experimental data reveal a $\Delta \tau^{f/b} =  \mathrm{7 \pm 2} as$ for (1S)-(-)-camphor ($\Delta \tau^{f/b} = \mathrm{- 7 \pm 2}$ as for (1R)-(+)-camphor) for the SB2 and a null $\Delta \tau^{f/b}$ for the higher energy SB3.

\begin{figure}[h]
\begin{center}
\includegraphics[width=0.8\textwidth]{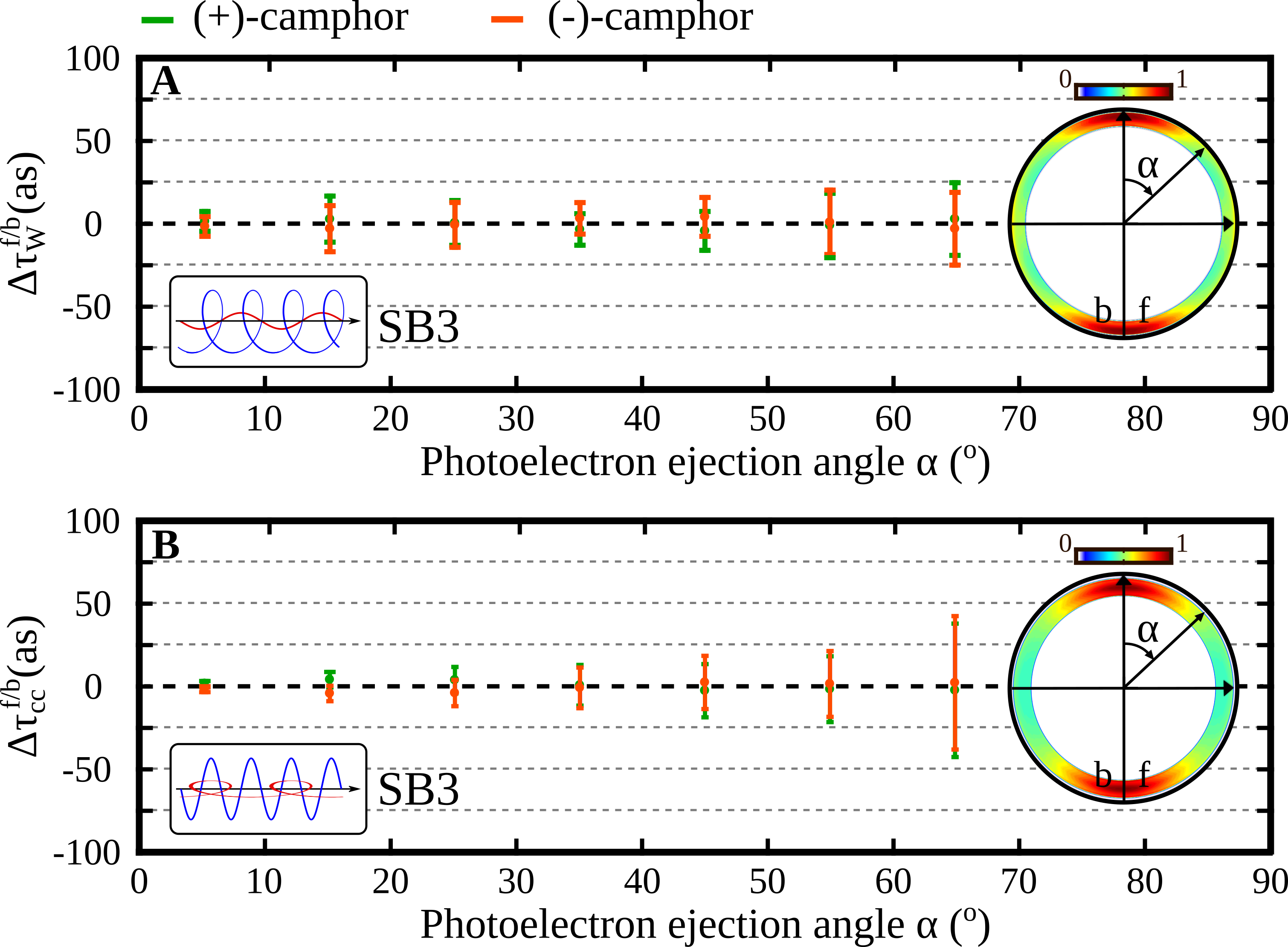}
\caption{Angle-resolved f/b differential delay for SB3. In (\textbf{A}) when using circularly polarized UV and linearly polarized IR and in (\textbf{B}) when using linearly polarized UV and circularly polarized IR.}
\label{Angular_SB3}
\end{center}
\end{figure}

We have performed an angular-resolved analysis of the differential photoionization delay of the SB3 by slices of 10$^\circ$ instead of by quadrant, in order to see if it vanishes for all ejection angles. Figure \ref{Angular_SB3}(A) shows the evolution of the differential Wigner delay with electron ejection angle for SB3. The differential delay remains 0 within the error bars for all ejection angles, which confirms the validity of our initial conclusion. One can notice that the error bars on this measurement are larger than those on SB2. This is caused by the low level of the signal, which is typically one order of magnitude lower for SB3 than for SB2. Repeating the measurements with higher accuracy (active stabilization of the delay line, higher repetition rate of the laser, longer pulses with narrower spectrum...) could enable us to reduce the error bars and reveal possible small but non-zero values. Figure \ref{Angular_SB3}(B) shows the evolution of the differential continuum-continuum delay with electron ejection angle for SB3, when the UV field is linearly polarized and the IR field is circularly polarized. The $\Delta \tau_{cc}^{f/b}$ also remains 0 within the error bars for all ejection angles.

\subsection{Sampling and phase measurements of the oscillating signals}

The temporal resolution of our experiment is determined by the accuracy with which we are able to measure the phase of the oscillating (sine) function. It depends on the way we sample the sine function as well as on the algorithm used to extract the phase. From the sampling point of view, the experimental acquisition time sets a limit and imposes the total number of samples that can be acquired in a reasonable time. In our experimental conditions we recorded 5 oscillations with 10 points per oscillations, and sequentially repeated the measurement for 5 consecutive delay scans in each enantiomer. The overall acquisition time was roughly 10 hours. Recording 5 oscillations fulfills the standard set by IEEE to properly sample a sine function \cite{sedlacem05}. 

For the phase measurement, we compared two methods: Discrete Fourier Transform and 4-parameter sine fitting. The method can significantly influence the accuracy of the results, as demonstrated for instance in \cite{sedlacem05}. The 4-parameter fitting followed the procedure recommended by IEEE (IEEE-STD-1057). 

\begin{figure}[h]
\begin{center}
\includegraphics[width=0.8\textwidth]{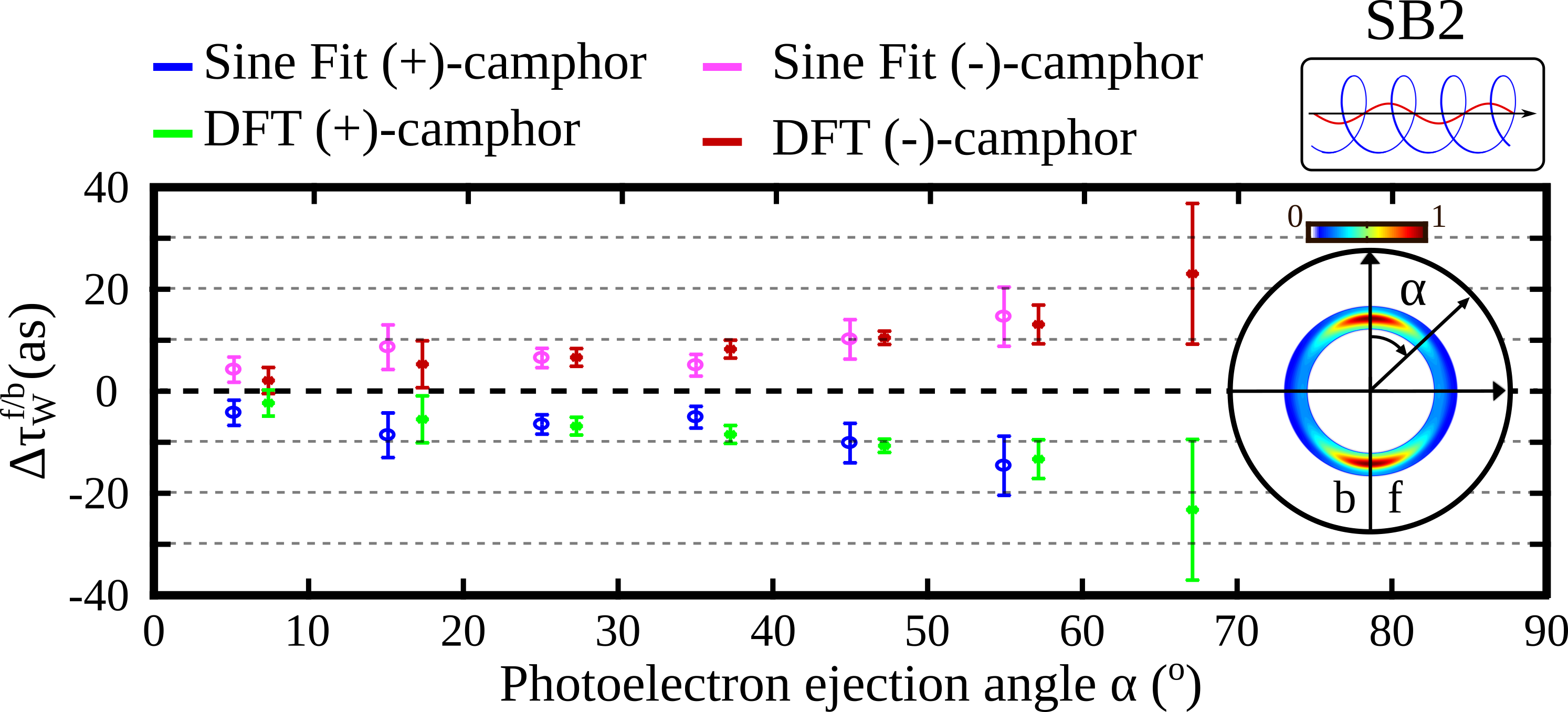}
\caption{Comparaison between Discrete Fourier Transform (DFT) and 4-parameter Sine Fit methods to extract the angle-resolved $\Delta \tau^{f/b}$ for sideband 2. An offset of 2 degree was added to the point extracted with the DFT method in order to clearly see the difference between the DFT and Sine Fit methods.}
\label{dft_vs_sine}
\end{center}
\end{figure}

Figure \ref{dft_vs_sine} shows the $\Delta \tau^{f/b}$ extracted from both methods as a function of electron ejection angle. The error bars are calculated by Student statistical analysis of the consecutive measurements. The results from the two methods are consistent within the error bars. The DFT provides smaller error bars, and we thus chose this procedure for our data analysis. 

\end{document}